\documentclass[emulateapj,twocolumn]{aastex63}
\usepackage{graphicx}
\usepackage{float}
\usepackage{amsmath}
\usepackage{subfigure}
\usepackage{natbib}
\usepackage{array}
\usepackage{gensymb}
\usepackage{float}
\restylefloat{table}
\newcommand{\head}[2]{\multicolumn{1}{>{\centering\arraybackslash}p{#1}}{\textbf{#2}}}
\usepackage{booktabs}
\usepackage{xcolor}

\begin{document}

\submitjournal{ApJ}
\accepted{July 24, 2021}

\title{Molecular gas filaments and fallback in the ram pressure stripped Coma spiral NGC 4921}
\author{Cramer, W. J.}
\affil{School of Earth and Space Exploration, Arizona State University, Tempe, AZ 85287, USA}
\author{Kenney, J.D.P.}
\affil{Department of Astronomy, Yale University, New Haven, CT 06511, USA}
\author{Tonnesen, S.}
\affil{Flatiron Institute, CCA, 162 5th Avenue, New York, NY 10010 USA}
\author{Smith, R.}
\affil{Korea Astronomy and Space Science Institute (KASI), 776 Daedeokdae-ro, Yuseong-gu, Daejeon 34055, Korea}
\author{Wong, T.}
\affil{Department of Astronomy, University of Illinois, Urbana, IL 61801, USA}
\author{J{\'a}chym, P.}
\affil{Astronomical Institute, Czech Academy of Sciences, Prague, Czech Republic}
\author{Cort{\'e}s, J. R.}
\affil{National Radio Astronomy Observatory Avenida Nueva Costanera 4091, Vitacura, Santiago, Chile}
\affil{Joint ALMA Observatory, Alonso de C{\'o}rdova 3107, Vitacura, Santiago, Chile}
\author{Cort{\'e}s, P. C.}
\affil{National Radio Astronomy Observatory Avenida Nueva Costanera 4091, Vitacura, Santiago, Chile}
\affil{Joint ALMA Observatory, Alonso de C{\'o}rdova 3107, Vitacura, Santiago, Chile}
\author{Wu, Y.-T.}
\affil{National Astronomical Observatory of Japan, Mitaka, Tokyo 181-8588, Japan}

\begin{abstract}

We investigate the effects of ram pressure on the molecular ISM in the disk of the Coma cluster galaxy NGC 4921, via high resolution CO observations. We present 6\arcsec \, resolution CARMA CO(1-0) observations of the full disk, and 0.4\arcsec \, resolution ALMA CO(2-1) observations of the leading quadrant, where ram pressure is strongest. We find evidence for compression of the dense interstellar medium (ISM) on the leading side, spatially correlated with intense star formation activity in this zone. We also detect molecular gas along kiloparsec-scale filaments of dust extending into the otherwise gas stripped zone of the galaxy, seen in HST images. We find the filaments are connected kinematically as well as spatially to the main gas ridge located downstream, consistent with cloud decoupling inhibited by magnetic binding, and inconsistent with a simulated filament formed via simple ablation. Furthermore, we find several clouds of molecular gas $\sim 1-3$ kpc beyond the main ring of CO that have velocities which are blueshifted by up to 50 km s$^{-1}$ with respect to the rotation curve of the galaxy. These are some of the only clouds we detect that do not have any visible dust extinction associated with them, suggesting that they are located behind the galaxy disk midplane and are falling back towards the galaxy. Simulations have long predicted that some gas removed from the galaxy disk will fall back during ram pressure stripping. This may be the first clear observational evidence of gas re-accretion in a ram pressure stripped galaxy.

\end{abstract}

\section{Introduction}

Ram pressure stripping is a key process by which gas is removed from galaxies in clusters, leading to star formation quenching, and the transformation of cluster galaxies from blue to red \citep{Gunn+72, Dressler+80, Whitmore+93, Poggianti+99}. The effects of ram pressure on the lower density, ionized and atomic gas in galaxies have been observed extensively. 

Recently, there have been many studies of spectacular kpc scale tails of gas being actively stripped from cluster galaxies \citep{Yagi+10, Fossati+12, Ebeling+14, Jachym+14, Poggianti+17, Boselli+18, Cramer+19}. In addition, significant amounts of molecular gas have been found in the extraplanar gas tails of ram pressure stripped galaxies \citep{Vollmer+08, Jachym+14, Jachym+17}, and in some cases studied at high resolution with ALMA \citep{Jachym+19, Moretti+19}. High resolution (on the scale of $\sim$100 pc) studies of the effects of ram pressure on the molecular gas \textbf{in disks} have been rarer. There is much to be learned about the effects of ram pressure on the gas in the disk.

Observations of molecular gas in the disks of ram pressure stripped galaxies have found compression of gas on the leading side \citep{Vollmer+12, Cramer+20}. Studies have also found evidence for enhanced star formation on the leading side in both observations \citep{Koopmann+04, Ebeling+19, Roberts+20} and simulations \citep{Cen+14, Steinhauser+16, Genina+19, Troncoso+20}. The effects of triggered star formation in ram pressure affected galaxies is strong enough that some analyses have found on average they tend to lie above the typical star formation rate to stellar mass relation \citep{Vulcani+18, Poggianti+20, Roberts+20}. Furthermore, studies have also found a much higher molecular gas mass to stellar mass fraction in ram pressure stripped galaxies than that observed in undisturbed galaxies \citep{Jachym+17, Moretti+19, Noble+19}. However, to study the evolution of the dense ISM that can lead to the observed enhanced gas fractions and the formation of compressed zones, high-resolution observations of relatively face-on galaxies are necessary. With these, we can address the following important questions: what is the structure and what are the properties of the ram pressure compressed gas? How much gas gets left behind in stripped zones, for how long, and what effect does this have on the subsequent evolution of stripped galaxies? Can observations of the dynamics of the gas in the disk in high resolution help us to explain the elevated incidence of AGN in ram pressure affected galaxies observed by \citet{Poggianti+17a}? Some simulations support that ram pressure stripping can drive gravitationally bound gas inward, accelerating black hole accretion rates and possibly triggering AGN \citep{Tonnesen+09, Ramos+18, Ricarte+20}.

Moreover, whether, and how much, ram pressure accelerated gas actually reaches the escape velocity and is stripped away into the ICM is unknown. If a stripped gas cloud does not reach escape velocity over a timescale of $\sim$1 orbit, it will likely fall back onto the host galaxy \citep{Koppen+18}. Gas clouds can fail to reach escape velocity if ram pressure decreases over time, and/or if gas moves into the shadow of the disk, as this gas will no longer be accelerated as strongly away from the host galaxy \citep{Tonnesen+10, Tonnesen+12, Jachym+13}. Some of the stripped gas may also cool and condense. In this case, if the gas has not yet reached the escape velocity, it will fall back onto the host and be re-accreted, either colliding with gas in the disk, or passing through and oscillating around the disk midplane \citep{Koppen+18}. Studies with simulations of gas stripping have found that a significant mass fraction of stripped gas is re-accreted by the host galaxy, in some cases up to 50\% \citep{Vollmer+01, Schulz+01, Roediger+05, Kapferer+09}. However, the impact of both the disk wind angle and time variable ram pressure on the re-accretion rate is important to consider, and results in significant variation in the amount of fallback in simulations \citep{Quilis+17, Koppen+18, Tonnesen+19}. While there has been some evidence found supporting the fallback of stars based on morphological analysis  \citep{Kenney+14}, and gas and star fallback has been seen to occur in simulations, clear observational evidence has not yet been found to compare with predictions from simulations.

Constraining the amount of fallback in ram pressure stripped galaxies is very important for a complete understanding of the effects ram pressure has on galaxy evolution. Re-accretion of stripped gas results in delayed quenching times, as re-accreted gas continues to fuel star formation, which can result in much less actual quenching than is predicted purely from the gas loss as calculated from the Gunn \& Gott formula \citep{Quilis+17}. Furthermore, clouds that fall back onto the galaxy result in a kinematic disturbance of the disk as they collide and some pass through to the other side. Stars formed in these clouds that fall back end up forming a thick disk component of younger stars with higher velocity dispersion than the surrounding disk \citep{Kapferer+09, Abramson+11, Steyrleithner+20}. Large enough clumps of stars falling back onto the disk may even form observable stellar streams of bluer stars in the outer galaxy \citep{Kenney+14, Cramer+19}. Learning how much fall back there is and where it impacts the disk is key to a proper understanding of the evolution of galaxy stellar populations and structure in affected galaxies. Identifying cases of stripped gas fallback with observations is an important first step in this process.

\subsection{NGC 4921}

\begin{figure*}
	\plotone{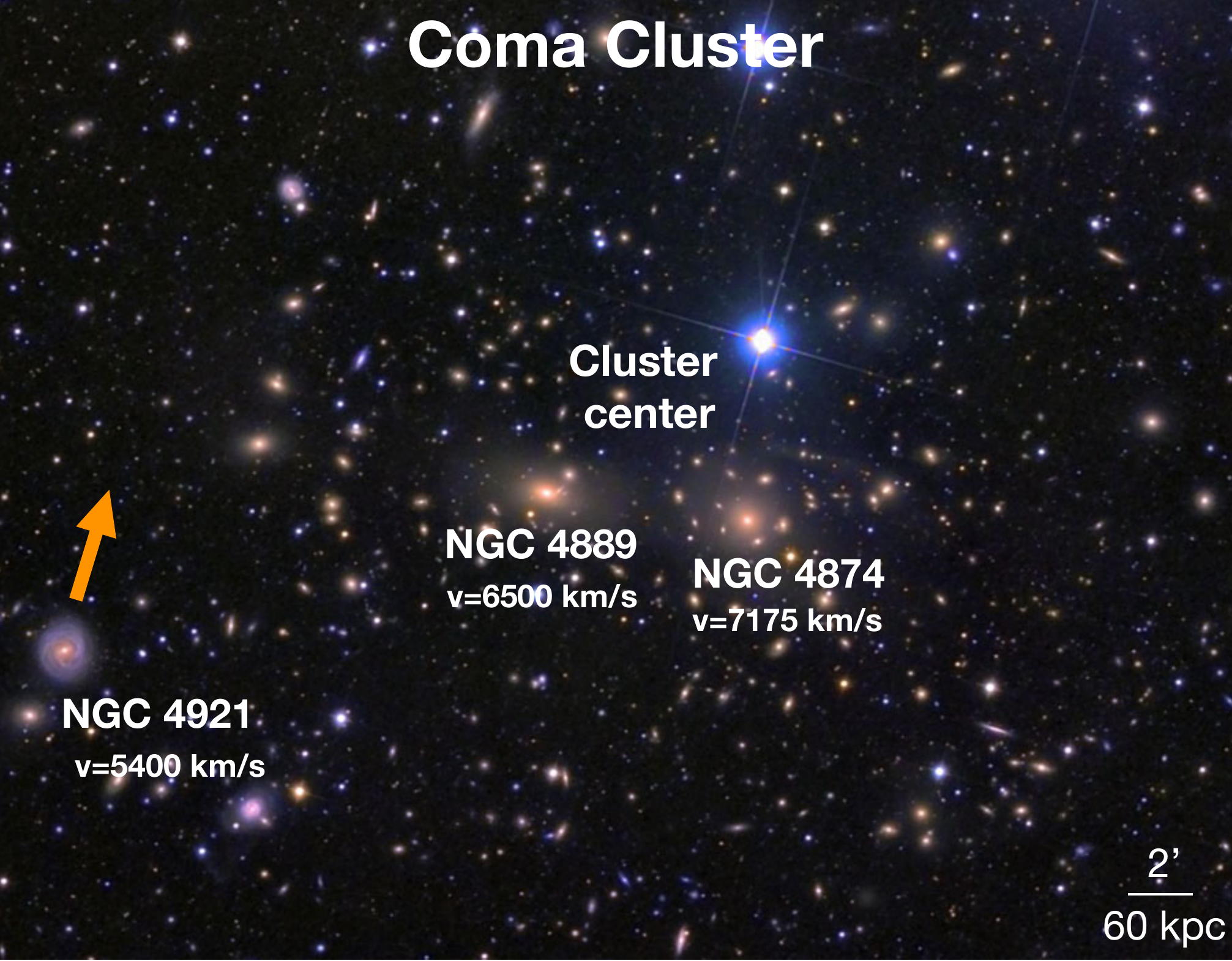}
	\caption{A false color image of the Coma cluster from an image taken by Dean Rowe with an STL11000 camera with Astrodon LRGB filters. We have marked the locations (as well as the line of sight velocities) of NGC 4921 and the largest ellipticals at the cluster center. The orange arrow indicates the projected direction of motion of NGC 4921 through the cluster, based on the morphology of its HI disk. Image credit: \url{http://deanrowe.net/astro/}.}
	\label{fig:Coma_full}
\end{figure*}

The Coma cluster is the nearest very massive cluster and thus, a great environment in which to study the effects of ram pressure at high resolution. There is clear evidence for the effects of strong ram pressure in Coma, most strikingly illustrated by the long tails of stripped gas found and studied in such works as \citet{Yagi+07, Yagi+10, Cramer+19, Chen+20}. An excellent galaxy in the Coma cluster in which to study the effects of RPS on the dense ISM in the disk is NGC 4921, a massive (M$_B$=-22), nearly face-on ($i \sim 13^{\circ}$) spiral. We show in Figure \ref{fig:Coma_full} the location of NGC 4921 in the cluster, as well as its projected velocity relative to cluster center ($\sim -1500$ km s$^{-1}$), and projected distance to cluster center (700 kpc). We estimate a significant component of the orbital motion of NGC 4921 in the cluster is azimuthal not radial, based on compressed dust and HI on the NW side, and slight HI elongation to the SE, although we note that the orientation of these features does not necessarily track the instantaneous direction of ram pressure, but may retain some signature of prior wind angles. The galaxy is also blueshifted by 1500 km s$^{-1}$ with respect to cluster center, so there is a component of the ram pressure wind acting perpendicular to the disk, and thus pushing gas in the redshifted direction to the observer. The outer galaxy HI asymmetry angle in NGC 4921 is PA$=345\degree$ (Kenney et al. (in prep)). Analysis of ram pressure stripping simulations shows that these angles are generally good indicators of the projected wind direction, to within $5-10\degree$ in cases of strong increasing ram pressure (Mas et al. (in prep)), like NGC 4921. The orientation of NGC 4921, nearly face-on, and its relative closeness makes it possible to study the effects of ram pressure on the ISM in the disk at extremely high resolution. Most studies of the ISM in ram pressure stripped disks have been of highly inclined galaxies, as this orientation is more common to the observer. 

NGC 4921 has a truncated HI disk with a head-tail morphology characteristic of galaxies undergoing ram pressure, with the largest degree of truncation on the northwest side of the galaxy, presumably the leading side \citep{Kenney+15}. The galaxy has been previously observed with HST in F350LP, F606W, and F814W, and the images are quite deep (proposal ID: 10842 \& 12476, PI: Cook).

\begin{table}
\centering
\caption[]{NGC 4921 Properties.}
\label{N4402}
\begin{tabular}{ll}
\hline
\hline
\noalign{\smallskip}
RA (J2000)        & $13^{\rm h}01^{\rm m}26^{\rm s}$.12\\
Dec (J2000)       & $+27^{\circ}53'09''.56$\\
type                   & SBab\\
$z$                   & 0.01829\\
length scale                    & 0.05''$=25$ pc\\
Projected distance                   & 700 kpc\\
from cluster center                        &  \\
$v_{\rm LSR}$          & 5400~km\,s$^{-1}$\\
$v_{\rm gal} - v_{\rm Coma}$          & -1500~km\,s$^{-1}$\\
M$_B$ mag  & $-22''$\\
PA, inclination & $177^{\circ}$, $13$\degree$^2$\\
%rotation velocity      & \\
\noalign{\smallskip}
\hline
\noalign{\smallskip}
\end{tabular}
\tablenotetext{2}{See Section \ref{modelling} for details}
\end{table}

Studies show that the ISM on the leading side of NGC 4921 has a very different morphology, as well as different star forming properties, than other regions of the disk. \citet{Lee+16} found that on the leading side of the galaxy, the star clusters formed were more massive and more luminous, possibly due to the formation of giant molecular clouds (GMCs) with unique properties due to ram pressure compression of gas. The dust structures on the leading side of the galaxy has also been studied in depth with HST \citep{Kenney+15}. They found thin, kpc long filaments of dust, extending out into the otherwise dust-stripped zone of the galaxy. Some of these filaments have luminous young star complexes at the heads. \citet{Kenney+15} suggested the structure of these filaments was consistent with magnetic binding inhibiting the decoupling of the ISM under ram pressure. The filaments appear similar to "elephant trunk" features in HII regions observed in the Milky Way (although the filaments in NGC 4921 are many times larger in scale), which have been suggested to be the result of the influence of magnetic fields \citep{Carlqvist+03}. \citet{Carlqvist+13} examined the filaments in NGC 4921, but did not include any discussion of the influence of ram pressure in NGC 4921, instead proposing the filaments may be magnetically supported spiral arm outgrowths. Other regions of the main ring of dust on the leading side appear to have been pushed in radially, creating "c-shaped" concavities \citep{Kenney+15}. The clarity of these unique structures in dust suggested that they may also be visible with sufficiently deep ALMA observations. Such deep, high resolution observations of molecular gas allows us to track the impact of RPS on the dense ISM in the disk.

\section{Observations}
\subsection{ALMA}

We observed the northwest quadrant (the leading side) of the galaxy NGC 4921 with ALMA in 2016-17 during Cycle 4, (project code 2016.1.00931.S, PI: Kenney). We observed in Band 6, centered around the CO(2-1) line at 230.36 GHz, in a 50'' $\times$ 20'' rectangular area. 

For the CO(2-1) data, the full mosaic is comprised of 7 pointings with the 12m array, each with a half power beam width (HPBW) of 25'', and 25 pointings with the 7m Atacama Compact Array (ACA), each with a HPBW of 44''. The velocity resolution of the data is 1.2 km s$^{-1}$ and the total time on source with the 12m array is 156 minutes, and with the ACA, 31 minutes.

Flux calibration is based on observations of Ganymede (ACA), and J1229+0203 (12m), bandpass calibration uses J1229+0203 (12m \& ACA), and the phase calibrator is J1303+2433 (12m) and J1331+3030 (ACA).

The data have been calibrated using the CASA pipeline, and combined using the CASA software package (version 5.4.0, \citet{McMullin+07}) with the \textit{tclean} command after checking for proper weighting. No significant continuum emission is found in the spectral windows around the CO(2-1) line. We chose to weight the data with a Briggs robustness of 2.0 (closest to natural weighting). The resulting beam size is  0.45'' $\times$ 0.35'' (220 $\times$ 170 pc).

The data were cleaned using the CASA \textit{tclean} procedure; due to extended diffuse structure in the data cube, we utilized the multi-scale clean option. Emission was found in 37 binned 5 km s$^{-1}$ channels, ranging from $5340-5525$ km s$^{-1}$.  The data were cleaned down to an RMS level of $\sim$ 6 mJy beam$^{-1}$. The final maps have some slight negative bowl (no greater than $\sim$ 2.5 sigma) features due to the lack of total power observations of the galaxy. We found that the default CASA \textit{immoment} routine made decent moment maps, but did not capture some of the low-level emission seen in the channel maps. We thus used a dilated masking approach implemented in python\footnote{\url{https://github.com/tonywong94/maskmoment}} to exclude noise before calculating the moment images. The mask is defined from an initial 4$\sigma$ contour and expanded to the enclosing 2$\sigma$ contour. From totaling the moment 0 map, we measured a total CO(2-1) flux (of all flux above 1.5 sigma significance) of $46 \pm 15$ Jy km s$^{-1}$ in the leading quadrant of the galaxy.

\subsection{CARMA}

As part of a survey of Coma cluster spirals (project c1025; PI T. Wong),
observations of CO(1--0) emission from NGC 4921 were conducted in 2012 August and October using the Combined Array for Research in Millimeter-Wave Astronomy (CARMA) in the most compact `E' configuration (baselines approximately 2.4--24 k$\lambda$) and the slightly less compact `D' configuration (baselines approximately 4--40 k$\lambda$).  Total on-source integration time was approximately 6.8 hours in E configuration and 5.5 hours in D configuration.  The galaxy was observed in a 7-point hexagonal mosaic with pointings separated by 30\arcsec\ (approximately half the primary beam width of the 10~m telescopes), yielding a half-power field-of-view (FOV) with radius $\approx$50\arcsec.   Two spectral windows of 191 channels each were allocated to the CO line, covering radial velocities from 5140--5800 km s$^{-1}$ with a channel spacing of 3.4 km s$^{-1}$.

The data were calibrated with the MIRIAD software package \citep{Sault+95}. Frequency-dependent (passband) gains were determined using observations of one of the bright quasars J1927+739, 3C273 or OJ 287 (J0854+201).  The planet Mars was used as the primary flux calibrator; for the D-array track a planet was not available and a flux of 4.5 Jy was adopted for OJ 287 based on calibrated measurements within a few days of the observation.  Antenna gain calibration was performed using observations every half hour of the quasar J1310+323.

The calibrated visibilities were imaged in MIRIAD and deconvolved using an implementation of SDI CLEAN designed for mosaics (task MOSSDI2).  CLEAN components were searched for down to the 2$\sigma$ level over the entire region where the sensitivity was within a factor of 3 of the FOV center.  Cubes were generated with 1\arcsec\ pixels and 10 km s$^{-1}$ wide channels across a velocity range of 5240 to 5590 km s$^{-1}$ (radio definition, LSR frame, which we use throughout this paper). The synthesized beam of the data cube was 6.5'' $\times$ 4.8'', with an RMS noise of 27 mK per channel. 

To create the moment maps of the CARMA data we again used a dilated masking approach\footnote{\url{https://github.com/tonywong94/idl\_mommaps}} to exclude noise before calculating the moment images. The mask is defined from an initial 4$\sigma$ contour and expanded to the enclosing 1.5$\sigma$ contour. From totaling the moment 0 map, we measured a total CO(1-0) surface brightness (of all emission above 2.5 sigma significance) of $53 \pm 12$ Jy km s$^{-1}$.

\section{Analysis}
\subsection{Molecular gas morphology from CARMA}
\label{CARMA_morph}

\begin{figure}
	\plotone{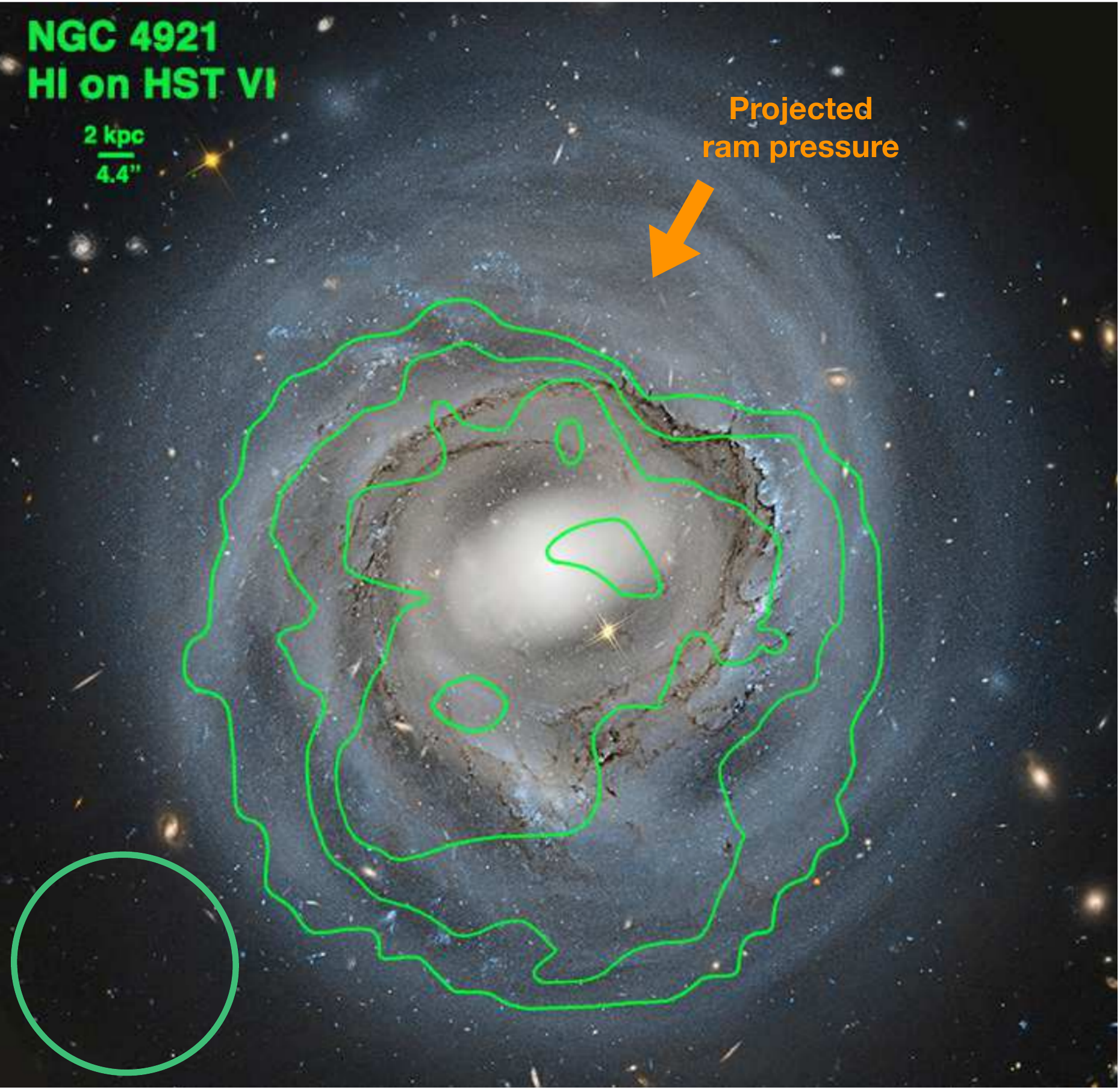}
	\caption{VLA HI intensity (moment 0) contours on HST color (F606W+F814W filters). HI contour levels vary by 1,3,5,7 times 17.5 Jy m s$^{-1}$ beam$^{-1}$. We note the asymmetric distribution of the HI, which is consistent with ram pressure acting from NW to SE. The beam is shown at the bottom left. This is based on a similar figure from \citet{Kenney+15}.}
	\label{fig:HI}
\end{figure}

\begin{figure*}
	\plotone{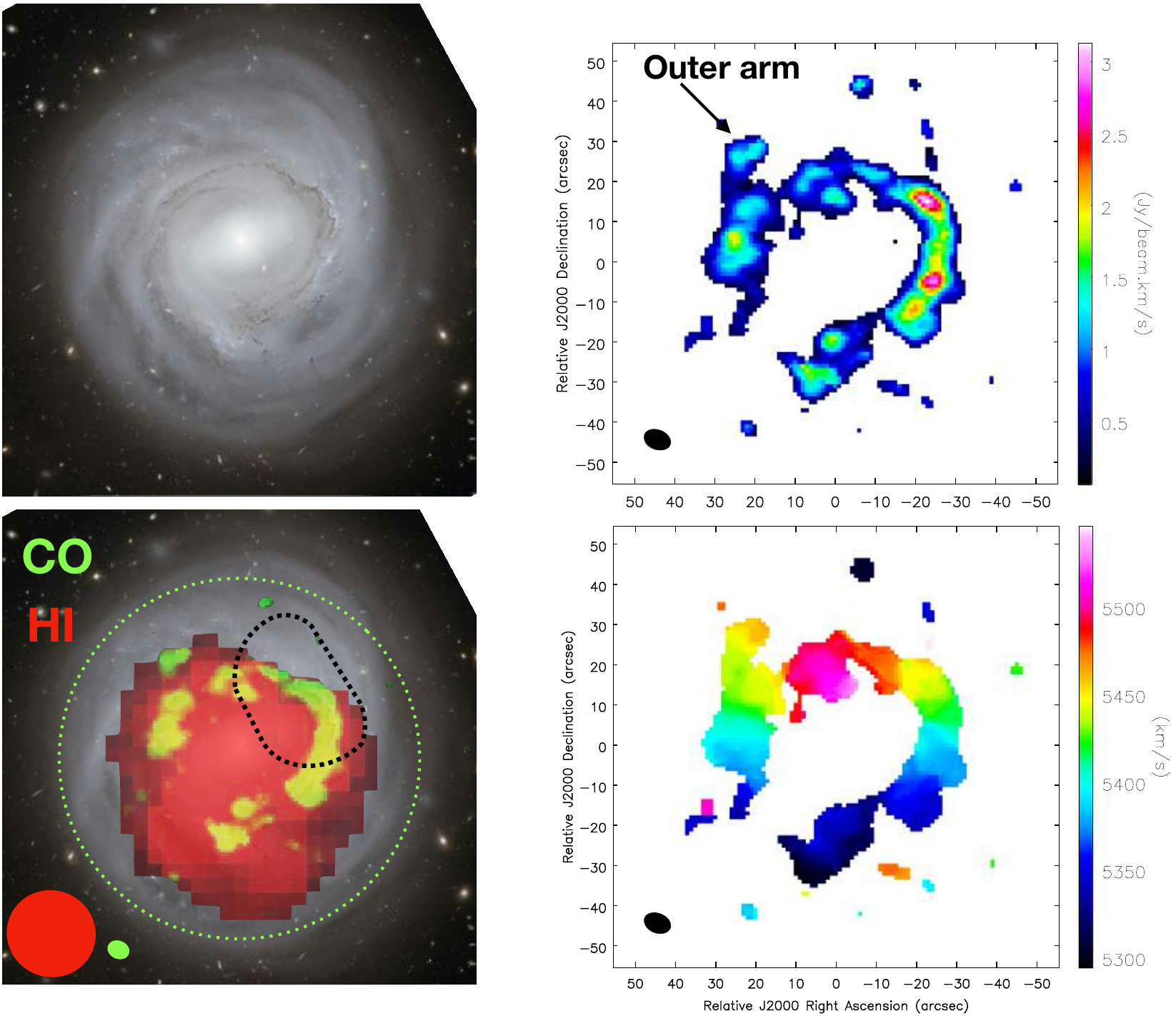}
	\caption{\textbf{Top Left}: An HST color image (F606W+814W) of the upper right side of NGC 4921 made with two filters, F606W in blue, F814W in redder, and an intermediate color added from averaging the two filters. \textbf{Top Right}: An intensity moment 0 map of CARMA CO(1-0) observations of NGC 4921. An outer arm of CO (also seen in dust) past the radius of the main dust ring is labeled. \textbf{Bottom Left}: CO intensity in green and HI intensity in red on the HST color image, to show the spatial coincidence between the CARMA CO(1-0), HI, and dust. The green dotted circle shows the field of view of the CARMA observations. The black dashed polygon shows the FOV of our ALMA observations. The HI and CO beams are shown at the bottom left. \textbf{Bottom Right}: A moment 1 velocity map of the CO(1-0).}
	\label{fig:CARMA_combo}
\end{figure*}

Similar to the dust distribution seen with HST (Figure \ref{fig:HI}), the CO(1-0), as observed with CARMA (see Figure \ref{fig:CARMA_combo}) is concentrated in a central ring, with a significant gap to the SE (the trailing side). The inner bound of the CO(1-0) detection ($r \sim 20''=12$ kpc) is just outside the stellar bar region of the galaxy, and the outer truncation radius of the CO(1-0) broken ring ($r \sim 30''=18$ kpc) is similar to the maximum detected radius of the HI. This broken ring feature has a total measured CO(1-0) flux of $\sim$ 45 Jy km s$^{-1}$ which accounts for 86\% of the total CO(1-0) flux detected in the galaxy with CARMA. The strongest peak in the CO(1-0) integrated flux ($\sim$ 85 mJy km s$^{-1}$) is on the leading side of the galaxy at an azimuthal angle of 30\degree \ west from north on the sky. This is near where we estimate that the galaxy is experiencing the strongest ram pressure based on the morphology of the HI (Figure \ref{fig:HI}), and the HST dust morphology \citep{Kenney+15}.

The molecular gas extent on the leading side is similar to the HI truncation radius as seen in Figure \ref{fig:CARMA_combo} bottom left. This suggests the atomic and the molecular gas may have both been pushed back, and compressed along the north side of the apparent molecular gas and dust ring (some atomic gas may have also been stripped and gravitationally unbound from the galaxy) and the remaining molecular gas is `effectively stripped` by disassociating into strippable density gas on a short timescale, as has been postulated by others, e.g. \citet{Kenney+04, Vollmer+12, Boselli+14}.

The distribution of the molecular gas as a function of azimuthal angle shows that a majority of the molecular gas is in the leading NW quadrant and the SW quadrant of the galaxy, which is rotationally downstream from the NW quadrant. The concentration of the molecular gas in the leading quadrant is coincident with the enhanced star formation in this same region observed by \citet{Lee+16}. About 180\degree \, opposite the estimated angle of incidence of ram pressure, the most shielded part of the galaxy, is a gap in the molecular gas ring. Outside the central broken ring, CO is also detected in the dust arm seen with HST on the NE side of the galaxy just past the dust ring at $r=32''$ (see Outer Arm in Figure \ref{fig:CARMA_combo} top right). Other `emission' blobs outside the main ring are likely to be residual noise peaks from the cleaning, because they are only detected in a single channel of the data cube. However, the blob to the N near the leading side, seen in black in the moment 1 map (at $\sim$ 5300 km s$^{-1}$) is detected in three channels of the cube at $\sim \, 3\sigma$, and its peak surface brightness is approximately three times higher than that of any of the other blobs outside the main ring. Its very offset velocity (up to $\sim$ 200 km s$^{-1}$) from the nearest ring CO is intriguing, but unfortunately this feature is not seen in any other wavelength we have access to, and is outside the field of view of our ALMA observations. If it is indeed a real feature it would be a likely candidate fallback feature; it will require further follow-up with deeper observations to confirm.

\subsection{Molecular gas distribution and morphology on the leading side observed with ALMA}

\begin{figure*}
	\plotone{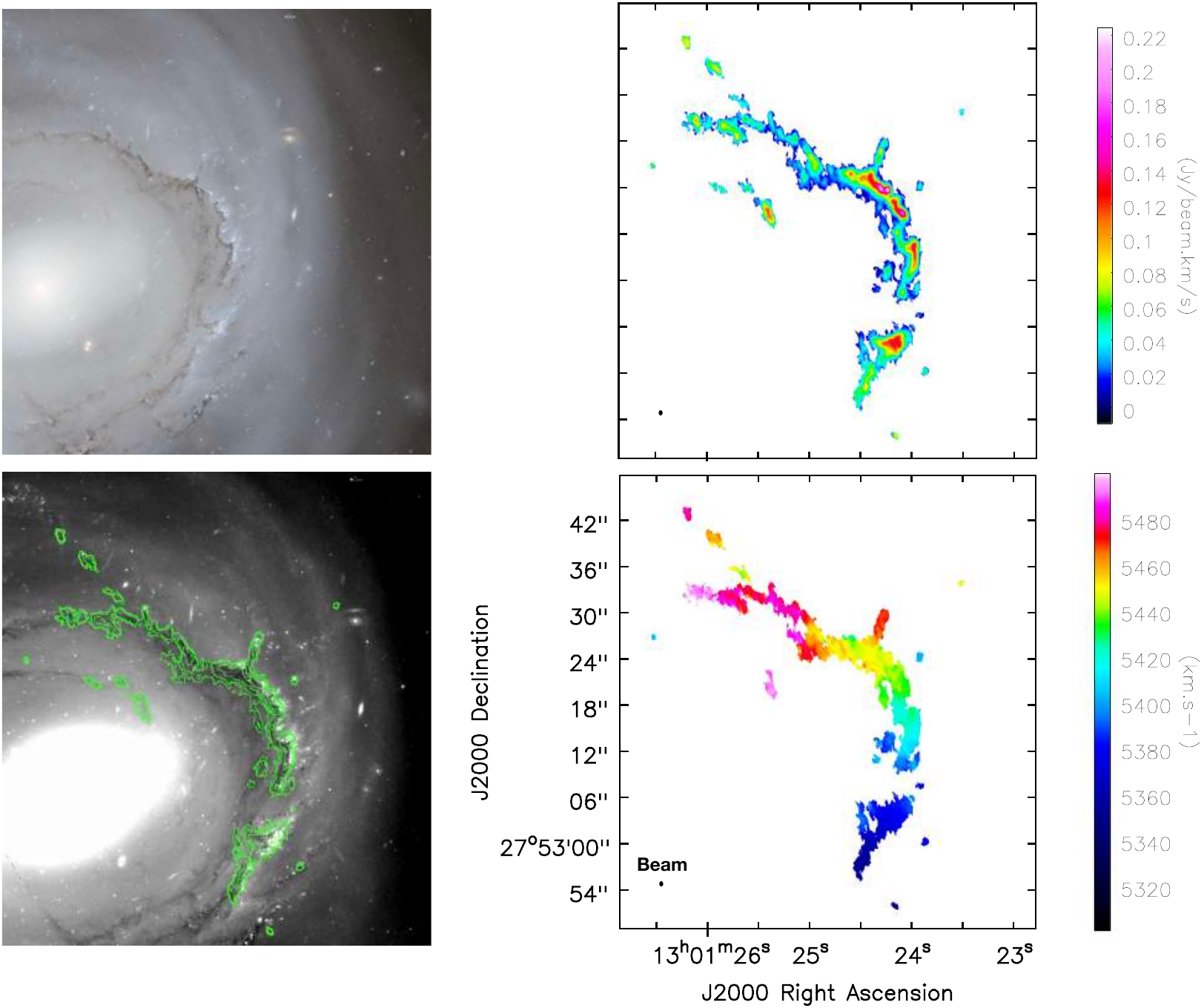}
	\caption{\textbf{Top Left}: An HST color image (F606W+814W) of the upper right side of NGC 4921 made with two filters, F606W in blue, F814W in redder, and an intermediate color added from averaging the two filters. The galaxy is rotating clockwise. \textbf{Top Right}:A moment 0 intensity map of the CO(2-1), where a flux cutoff is instituted by weighting by the sensitivity of the mosaic in order to display less noise. \textbf{Bottom Left}: CO intensity on F606W  with the contour levels ranging from $0.02 - 0.32$ Jy beam$^{-1}$ km s$^{-1}$, increasing by factors of two between levels. \textbf{Bottom Right}: An average velocity (moment 1) map of our ALMA CO(2-1) data.}
	\label{fig:moment_combo_weight}
\end{figure*}

\begin{figure*}
	\plotone{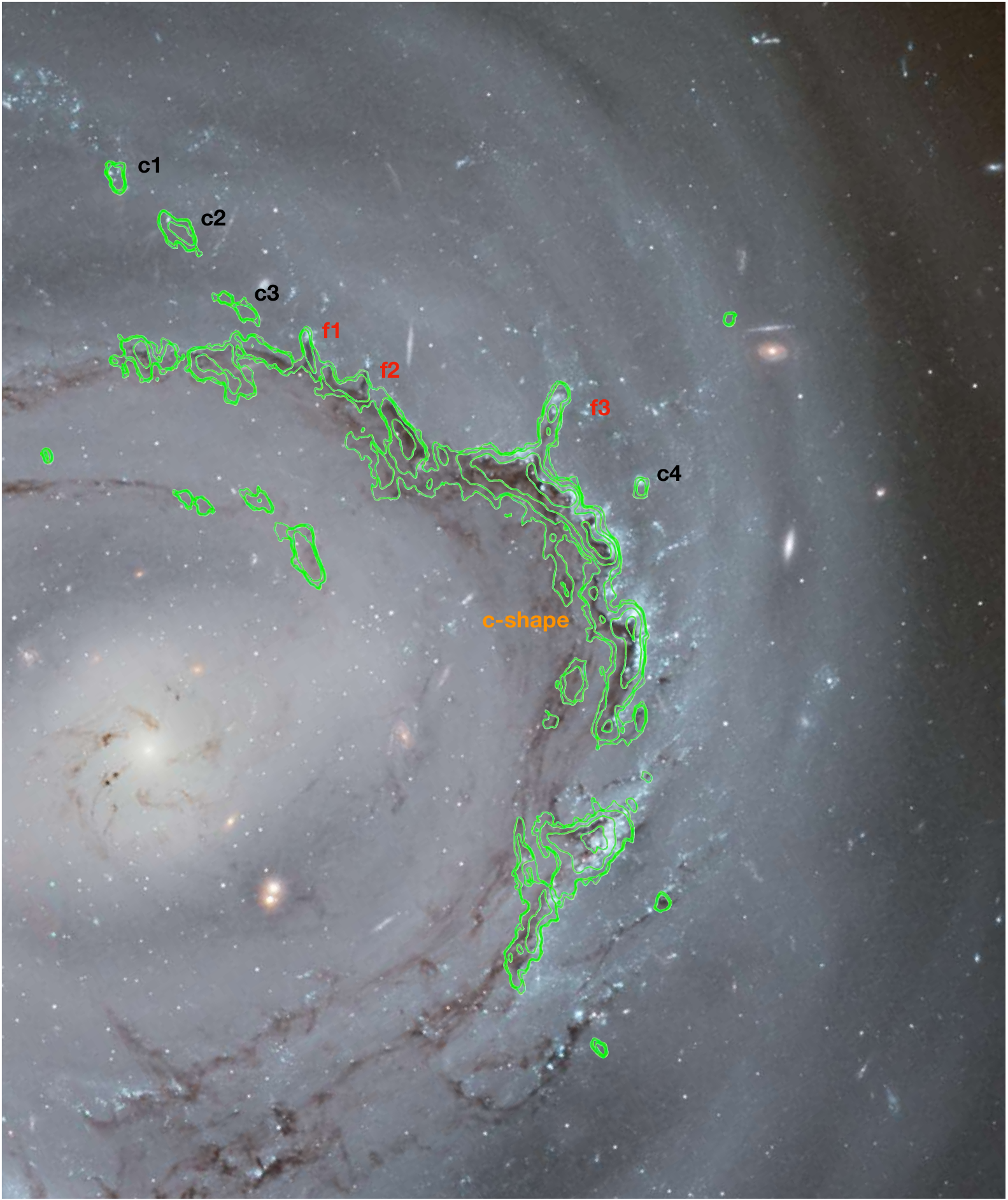}
	\caption{An overlay of the ALMA moment 0 map contours, with the same contour levels as those shown in Figure \ref{fig:moment_combo_weight}, on an HST color image (F606W+814W) reprocessed to show more of the dust structure (copyright: NASA/ESA/Kem Cook (LLNL), Leo Shatz). Four clouds of CO located beyond the main CO ring have been marked as c1, c2, c3, and c4. Prominent filaments we detect in CO are marked as f1, f2, and f3. The most well-resolved c-shape feature is also marked.}
	\label{fig:highres_zoom}
\end{figure*}
 
 Our ALMA observations of the leading side of NGC 4921, upon which ram pressure is most strongly incident, show the detailed morphology of the molecular gas, as traced by CO(2-1) in this region. We display our ALMA observations as moment maps where we have down-weighted pixels with low primary beam sensitivity so as to limit the noise, and make robust detections more apparent (Figure \ref{fig:moment_combo_weight}). We also show a larger version of this impressive, high resolution overlay in Figure \ref{fig:highres_zoom}. We see a prominent outer ridge of CO(2-1) at about $r \sim 25-28$'' from the nucleus, coincident with the region of dust and young star clusters seen in the HST image (Figure \ref{fig:highres_zoom}). This is also part of the broken ring feature seen with CARMA discussed in Section \ref{CARMA_morph}. The outermost front of the ridge of molecular gas contains the peak CO(2-1) line luminosity in the moment 0 map (Figure \ref{fig:moment_combo_weight}) of $\sim$ 2.5 $\times$ 10$^4$ K km s$^{-1}$ pc$^2$. We also detect part of an inner ring feature at $r\sim 22$'' in CO(2-1) that is seen in dust (Figure \ref{fig:highres_zoom}), although this feature is not detected as strongly in CO(2-1) as the main ring. Finally, we also detect four clouds past the radius of the main ring at $r=30$'' (marked as `c1', `c2', `c3', and `c4', in Figure \ref{fig:highres_zoom}) that appear to be decoupled from any surrounding molecular gas; the properties of these clouds are discussed later in this paper.
 
 Along the outer ring of molecular gas, we see similar structure when comparing the CO(2-1) and the dust (as seen via HST dust extinction). C-shaped pockets along the ring, also identified in \citet{Kenney+15}, are seen in the molecular gas. We also detect in CO(2-1) what appear to be three separate filaments of dust protruding from the main gas ring, indicated in Figure \ref{fig:highres_zoom} as `f1', `f2', and `f3' (although we note we detect CO only at the base of `f2', unlike `f1' and `f3' which are detected along their full length). This type of filament is not seen in dust in any other quadrant of the galaxy, suggesting that either they formed as a result of a recent, significant change in the strength of ram pressure, or that they do not persist as they rotate out of the leading quadrant. Filaments 1 \& 2 are the smallest, about 1.5'' (750 pc) in length, and both have bright blue star concentrations at the head. They have total masses of molecular gas along their length of $8.8 \times 10^6$ and $2.7 \times 10^6$ M$_{\odot}$ respectively. Filament 3 is the largest filament we detect, with a total length of 4.5'' (2.1 kpc) and a total mass of molecular gas of $4.3 \times 10^7$ M$_{\odot}$, and again, a clump of blue stars near the head, with a visible cap of dust at the top end. We show a list of the filament properties in Table \ref{tab:filament_table}, and find their average molecular gas surface density ranges from $\sim 25-45$ M$_{\odot}$ pc$^{-2}$. A similar filament of dust sticking out into the otherwise stripped zone, with a clump of young stars at the head, was also observed in the Virgo cluster galaxy NGC 4402, and was found to have an average surface density of only $7$ M$_{\odot}$ pc$^{2}$ \citep{Cramer+20}. We analyze and discuss these filaments further in Section \ref{filament}.
 
 Beyond the galactocentric radius of the main ridge of CO, there is no molecular gas other than that associated with the filaments and the previously mentioned decoupled clouds.

 \subsection{Molecular gas kinematics observed with ALMA}
\label{modelling}

Our high resolution ALMA observations allow us to study in detail the kinematics of the molecular gas on the leading side of NGC 4921, and assess how they may be influenced by ram pressure. However, to identify kinematics that deviate from normal motion associated with rotation, we must measure the average rotation curve of the galaxy, and generate a model of the galaxy velocity field based on it.

To generate a model of the average circular velocity in NGC 4921, we fit the moment 1 CARMA map (shown in Figure \ref{fig:CARMA_combo}) with DiskFit. DiskFit is a program that uses a tilted ring model to fit the kinematics of disk galaxies and is able to account for the beam size of the observations (see \cite{Sellwood+15} for documentation). Along with the moment 1 map, we also input a central $x$ and $y$ position of the galaxy from the peak of the surface brightness in the HST F606W image as well as an estimated central velocity $V_{\mathrm{sys}}$ based on that estimated from the HI velocity field. We also input a position angle (PA$=-3$\degree) and a disk ellipticity (0.1), based on the value determined from an isophotal analysis of the galaxy halo from a deep $r$-band image taken with the Dragonfly telescope (provided by Allison Merritt). We measured the isophotal properties of NGC 4921 from $r=20-64$'', and found that beyond a radius of $r=42$'' we reached a consistent measurement of the position angle and ellipticity. Within $r=42$'', irregular structure due to spiral arms, the bar, and bright star forming regions,
make values of the position angle and inclination more inconsistent. To further constrain our estimates we used the DiskFit bootstrapping procedure, whereby these parameters were varied until a minimized chi-squared value for the model fit was achieved. We find the best fitting values to be $\mathrm{PA} = 2 \pm 3$\degree, $e=0.06 \pm 0.04$ (corresponding to an inclination of $20^{+5}_{-9}$ degrees), and $V_{\mathrm{sys}} = 5400 \pm 5$ km s$^{-1}$.

\subsubsection{Measuring the systemic velocity}

The best fitting position angle based on the kinematic data matches very well with that measured from the optical image. The systemic velocity measured from the CARMA CO observations is quite different from that estimated from the HI data (5460 km s$^{-1}$); given the much lower relative resolution of the HI data, we regard the $V_{\mathrm{sys}}$ measured from our CARMA observations as more reliable. However, while the HI data does have lower resolution, the data are good enough to estimate the central velocity to better than $\sim$20 km s$^{-1}$. The HI velocity field shows a well-organized pattern, and the isovelocity contour increment for the moment 1 map given in \citet{Kenney+15} is 20 km s$^{-1}$. Since the moment 1 map shows a well-organized pattern, this implies that the local uncertainty in HI velocity is less than 20 km s$^{-1}$. If the uncertainty were larger, the isovelocity pattern would be noisy. This strongly supports that the 60 km s$^{-1}$ difference between HI and CO is real. 

The stripping direction is redshifted, and the HI is redshifted with respect to the CO, so this offset is likely due to ram pressure. The HI isovelocity pattern shows a modified spider diagram. There is a classic spider pattern present, but it is modified in that the outermost contours are ‘bent down’, meaning redshifted relative to pure rotation \citep{Kenney+15}. This kinematic pattern is consistent with influence from ram pressure affecting the kinematics of the outer galaxy; modelling published in \citet{Haan+14} found similar results. However, this new result (a 60 km s$^{-1}$ offset between HI and CO) strongly suggests that the mean HI velocities are systematically redshifted by ram pressure when compared to the denser molecular gas over the whole face of the galaxy. As one would expect in the case of ram pressure stripping, the outer region is more redshifted than the inner region, and the denser (molecular) gas is less affected than the lighter atomic gas. A similar bulk velocity offset between different gas phases was found in the ram pressure stripped galaxy J0201 (in this case between HI and H$\alpha$) \citep{Ramatsoku+20}.

\subsubsection{Estimating the inclination}

\begin{figure}
	\plotone{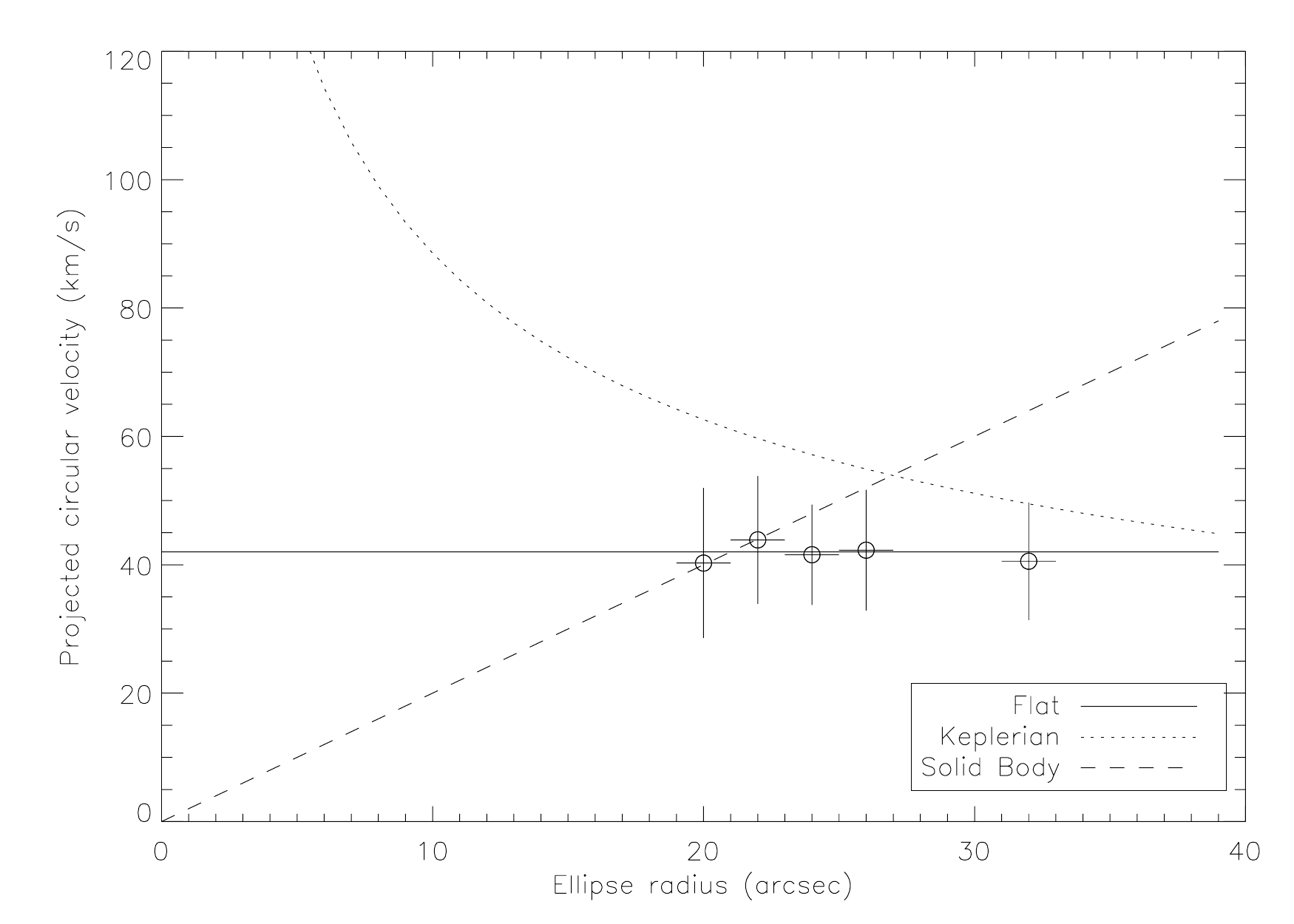}
	\caption{The best fitting projected rotation curve based on estimates with DiskFit. The plotted circular velocities have been projected such that they match the data by accounting for the inclination of the galaxy ($i=13$\degree). We find a flat rotation curve at the radius where we detect CO is the best fitting rotation curve. We have also plotted the maximum plausible rising rotation curve based on solid body rotation where $v \propto R$, and the maximum plausible falling rotation curve based on Keplerian decline where $v \propto R^{-1/2}$.}
	\label{fig:rotation_curve}
\end{figure}

The best fitting estimate of the inclination is similar to that measured from the Dragonfly image, although the uncertainty is large, which is not unexpected given that measuring the inclination in almost face-on galaxies from kinematics or isophotes is difficult without extremely high S/N data. We find that at the radius where we detect CO with our CARMA observations ($r=15-30$''), DiskFit finds a nearly flat rotation curve to be the best fit to the data, with a projected circular velocity of $62\pm$14 km s$^{-1}$ (Figure \ref{fig:rotation_curve}). At an inclination angle of $20\degree$, this would correspond to a circular velocity of $180\pm40$ km s$^{-1}$. However, given the luminosity of the galaxy (M$_B = -22$), one would expect from the Tully-Fisher relation that it should have a maximum circular velocity of $280\pm60$ km s$^{-1}$ \citep{Bohm+16}. For this to be true, the actual inclination angle of the galaxy would need to be $\sim$ $13^{+3}_{-2}$\degree, which is still within the 1$\sigma$ error range for the inclination estimated by the DiskFit fitting. Due to the likely higher reliability of the Tully-Fisher relation over kinematics or isophotes for estimating the inclination angle in close to face-on galaxies, we adopt an inclination angle of $13$\degree \, for the disk of NGC 4921 in this study, and rerun DiskFit using this value without allowing for variation. After rerunning with fixed inclination, the final DiskFit outputs are  $\mathrm{PA} = -1 \pm 3$\degree, and $V_{\mathrm{sys}} = 5402 \pm 5$ km s$^{-1}$.

\subsubsection{Comparing the model and data}

\begin{figure*}
	\plotone{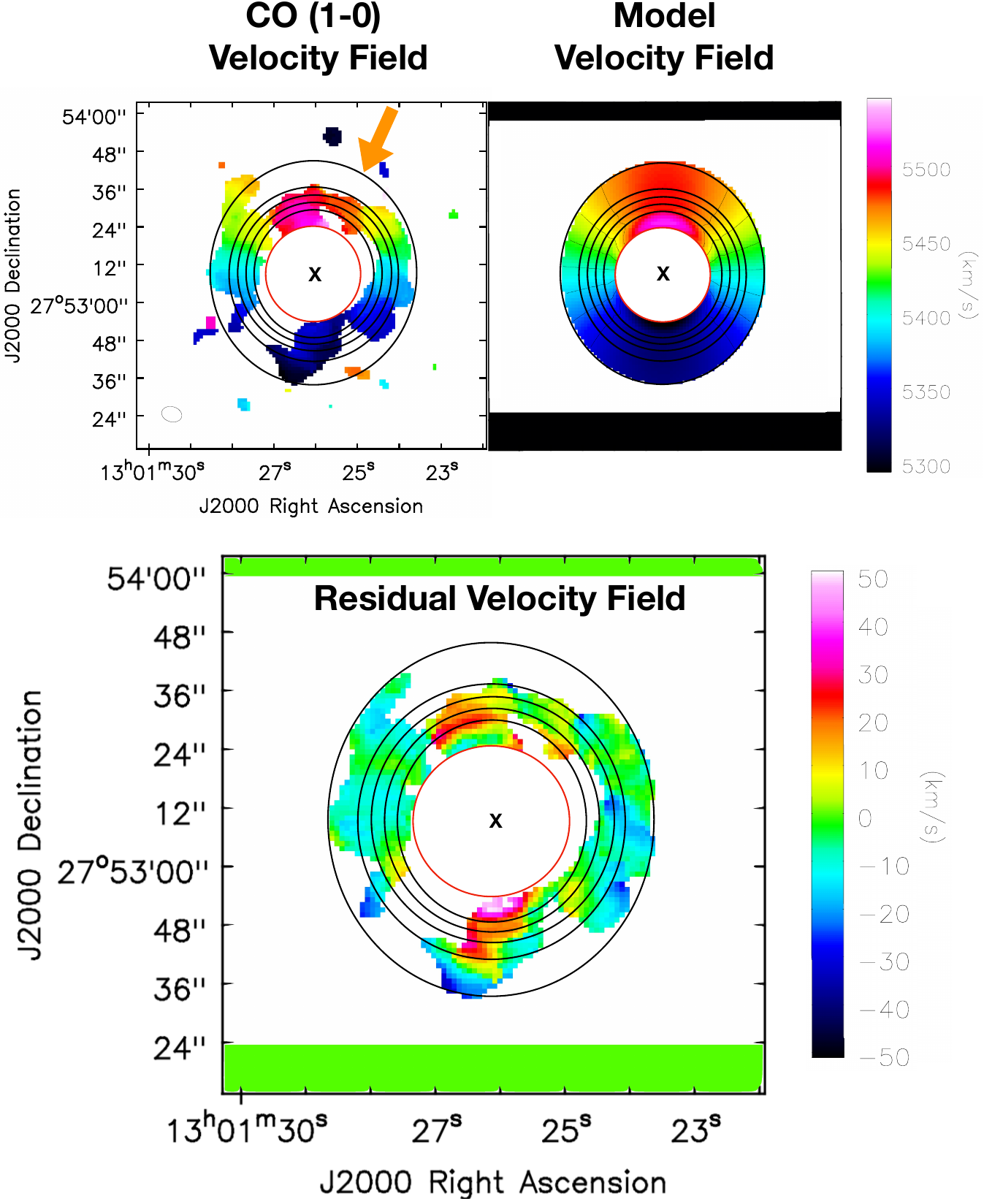}
	\caption{\textbf{Top Left:} The CARMA moment 1 map, and the elliptical fitting areas we chose for DiskFit to determine the best fitting rotation curve. The `x' indicates the optical center of the galaxy. The red ellipse shows the inner bound of the area we fit. The orange arrow indicates the projected ram pressure wind direction. \textbf{Top Right:} The resulting moment 1 map model of the galaxy generated from the rotation curve determined by DiskFit. Within the boundary set by the inner red ellipse Diskfit does no fitting and simply extrapolates the rotation curve from the outer parts of the galaxy inward. \textbf{Bottom:} The residual from subtracting the CARMA moment 1 map from the DiskFit model moment 1 map.}
	\label{fig:CARMA_residual}
\end{figure*}

The input image, resulting model, and residual map are shown in Figure \ref{fig:CARMA_residual}. The model does not agree in all regions of the galaxy equally well. To the north and south there are still significant velocity residuals of up to $\pm 50$ km s$^{-1}$ (Figure \ref{fig:CARMA_residual}). These residuals are especially large when considering that the projected maximum circular velocity is comparable in magnitude at $42\pm$8 km s$^{-1}$. Due to the fact that the rest of the CO ring has velocity residuals near $0\pm10$ km s$^{-1}$, and that we are confident in the best fitting inclination and position angle from DiskFit, we suspect these large residuals are probably due to a real kinematic disturbance in these regions. One possible source of non-circular motions may be the prominent bar feature in the galaxy having disturbed some gas within $r<20''$ (the approximate extent of the stellar bar) as the bar previously passed through. However, the regions with large residuals north and south along the major axis extend out to as far as $r=32''$. It is not obvious what the source of these residuals could be. We note that the region to the south contains a large concentration of optically blue stars, that are likely recently formed; the dust in this region also appears more irregular and disturbed. This region may have been previously compressed and kinematically disturbed by ram pressure. We would also expect the north side to be experiencing stronger ram pressure than the south side, as is also evidenced by the HI morphology. Along the quadrant of the galaxy in which we also have ALMA observations, the residuals from the CARMA data are between $0 \pm 10$ km s$^{-1}$, which given the CARMA channel width of 10 km s$^{-1}$, indicates a good match.

\begin{figure*}
	\plotone{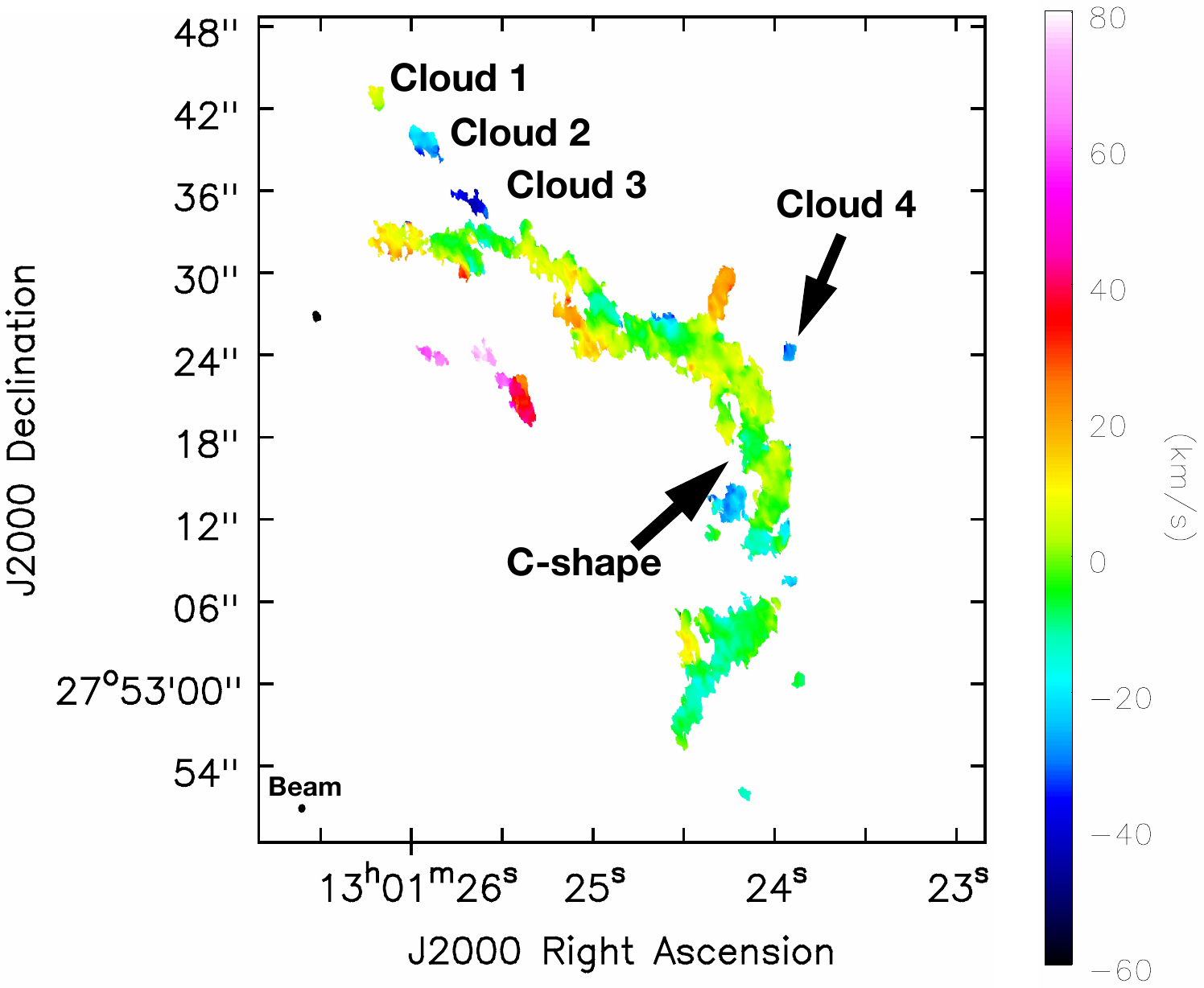}
	\caption{A residual map showing the ALMA moment 1 map subtracted from the DiskFit model of the circular velocity. The ALMA channel map width is 5 km s$^{-1}$, so residuals within $\pm$5 km s$^{-1}$ indicate that the velocity of the data is very close to the estimated circular velocity from our model. Features outside the main gas ring with significantly blueshifted velocity (falling back towards the galaxy) are identified as clouds $2-4$.}
	\label{fig:model_residual}
\end{figure*}

We interpolate the model image to the resolution of our ALMA data and compare to measure the kinematic residuals, where $V_{\text {residual}}=V_{\text {obs}}-V_{\text {model}}$, with the map shown in Figure \ref{fig:model_residual}. We find the main CO ring does not appear to show any widespread detectable offset from the predicted circular velocity in the direction of ram pressure. This is somewhat surprising as we might expect that the leading side, being exposed to strongest ram pressure, would have higher non-circular motions consistent with ram pressure pushing when compared to gas that had just begun to rotate into the leading side pressure. In the Virgo spiral NGC 4402 a similar region of strong, active star formation on the leading side of the galaxy had velocity offset in the direction of ram pressure of $\sim$20 km s$^{-1}$ \citep{Cramer+20}. We still do not have an explanation for the lack of a similar detectable kinematic offset in NGC 4921, although there are a number of factors which could cause the two regions to have different kinematics, including systematic such as uncertainty in the kinematic modelling, and physical such as gas density, the relative ram pressure and restoring force, and the disk-wind angle. We go into further detail on this in section \ref{RPS_estimate}, where we estimate the strength of ram pressure NGC 4921 is experiencing.

Examining the kinematic residuals from the ALMA data we find, like the CARMA data, the main ring of CO has only small deviations from the circular velocity predicted from our model. However, with the increased resolution of ALMA, we find significant velocity deviations from circularity in some of the already morphologically identified features of interest: decoupled clouds, filaments, and c-shapes. To interpret the meaning of these residuals, we must consider the that the velocity residual along the line-of-sight $V_{\text {residual}}$ is composed of some combination of three velocity components in the frame of reference of the galaxy: $V_{\text {z }}$ the vertical motion, $V_{t}$ the azimuthal motion, and $V_{r}$ the radial motion, described in equation 1

\begin{equation}
    V_{\text {residual}}=V_{\text {z }}\cos i+\sin i\left[V_{t} \cos \theta+V_{r} \sin \theta\right]
\label{velocity_eq}
\end{equation}

where $i$ is the inclination (which we estimate to be $13^{+3}_{-2}$\degree), and $\theta$ is the azimuthal angle with respect to the position angle of the major axis. 

\begin{figure*}
	\plotone{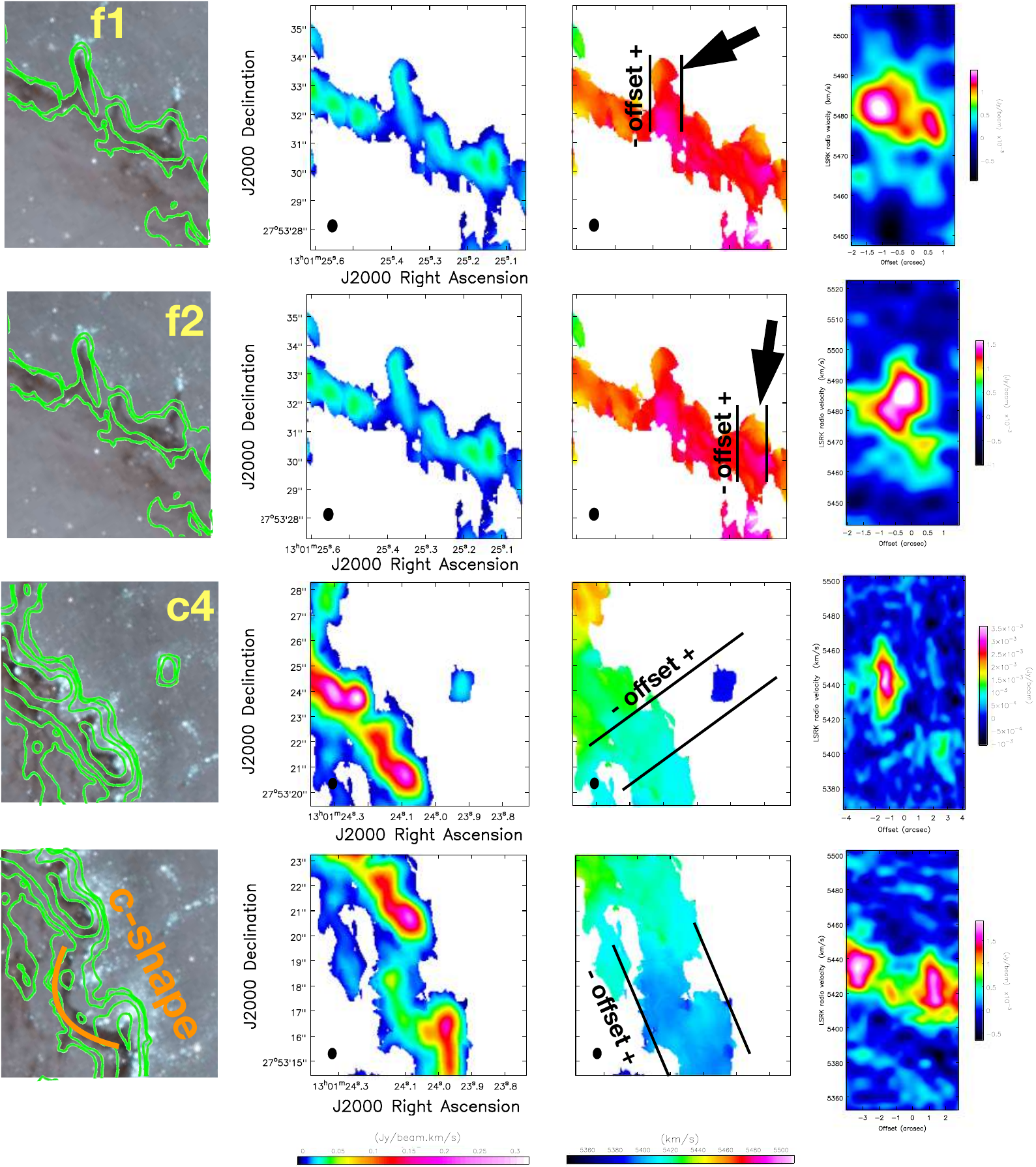}
	\caption{\textbf{Left}: A moment 0 intensity map of the zoomed-in region around the feature indicated with its label from Figure \ref{fig:highres_zoom}. \textbf{Middle Left} A moment 0 intensity map of the zoomed-in region. The beam is shown in the bottom left corner. \textbf{Middle Right}: A moment 1 velocity map of the zoomed-in region. \textbf{Right}: A PVD along the length of the feature shown on the left, with the offset direction indicated. Note that the velocity axis in f1 and f2 is more zoomed in, as the difference in velocity along the feature is so small. Cloud 4 is also plotted to show that it is completely decoupled from the surrounding disk, and is likely not the remnant of an evolved filament. Filament 3 is shown later in a separate figure (Figure \ref{fig:big_fil}).}
	\label{fig:filament_pvds}
\end{figure*}

The largest, and brightest in total flux clouds to the north outside the main CO ring are indicated with arrows in Figure $\ref{fig:model_residual}$ and labeled clouds 2 \& 3. They are blueshifted by $\sim 20-50$ km s$^{-1}$. Cloud 1 is further beyond the CO main ring, but does not show any velocity residual, and is in a region of the observations with low sensitivity. The more downstream large bright cloud, `Cloud 4', is also blueshifted by $20$ km s$^{-1}$, and we show a PVD of the cloud and nearby ring region showing they are kinematically separate in Figure \ref{fig:filament_pvds}. Clouds 2 \& 3 are close to the major axis, where $\theta \approx 0$, such that radial motions in the disk plane produce no component along the line of sight. However, from these observations alone we cannot determine the relative contributions to the observed blueshift due to vertical motion toward the galaxy disk, or azimuthal motion from these clouds rotating slower than the predicted circular velocity. If the motion of the clouds were entirely vertical, the correction for projection effects is very small as the disk is nearly face on, so the observed residual of $-20$ to $-50$ km s$^{-1}$ would correspond to a vertical streaming motion towards the observer of $-20$ to $-51$ km s$^{-1}$. If instead the motion were entirely azimuthal, the correction for inclination effects is large due to the inclination of the disk. Correcting for the inclination yields residuals due to azimuthal streaming of $-95$ to $-240$ km s$^{-1}$. Determining the likely components of the motion of these clouds requires additional evidence to consider outside of just our ALMA observations, which we explore further in Section \ref{fallback}.

The large filament (filament 3) is redshifted by $\sim$20 km s$^{-1}$, which may indicate the filament gas moving away from the observer. However, as the filament is located in between the minor and major axis, it may also be an azimuthal, or a radial motion, of up to 95 km s$^{-1}$. We discuss this further in Section \ref{filament}. The smaller filaments more towards the north, filaments 1 and 2, do not show a measurable deviation from normal rotation within our velocity resolution. Moment maps and PVD's of these features, as well as a c-shaped gap in the dust and gas ring discussed further in section \ref{c-shape}, can be seen in Figure \ref{fig:filament_pvds}.

\begin{table*}[hb]
\centering
\movetableright=-0.5in
\begin{tabular}{cccccc}
\hline
\head{1.5cm}{\textbf{Feature}} & \head{2.0cm}{\textbf{Integrated Flux (Jy km s$^{-1}$)}} & \head{2.0cm}{\textbf{Mass M$_{\mathrm{H_2}+\mathrm{He}}$ (M$_{\odot}$)}} & \head{1.5cm}{\textbf{Surface Area (pc$^2$)}} & \head{2.0cm}{\textbf{Surface Density (M$_{\odot}$ pc$^{-2}$)}} \\ \hline
\textbf{Filament 1} & 0.28 & 8.8 $\times$ 10$^6$ & 3.4 $\times$ 10$^5$ & 26 \\
\textbf{Filament 2} & 0.09 & 2.7 $\times$ 10$^6$ & 1.0 $\times$ 10$^5$ & 27 \\
\textbf{Filament 3} & 1.36 & 4.3 $\times$ 10$^7$ & 9.6 $\times$ 10$^5$ & 44 \\
\end{tabular}
\caption{\textcolor{black}{Properties of select CO features in NGC 4921. From left to right, in column (1): the feature name (abbreviated in some figures as `f1', `f2', and `f3'). (2): The integrated CO(2-1) flux based on the moment 0 map. (3): Total mass of H$_2$ \& He assuming $\alpha=4.3 \, \mathrm{M}_{\odot} \,\mathrm{pc}^{-2} \, (\mathrm{K} \, \mathrm{km} \, \mathrm{s}^{-1})^{-1}$ and $R_{21} \approx 0.8$, and the mass fraction of He/H$_2$ to be 1.34. (4) Surface area of features based on the spatial extent of the CO feature in dust. (5): Estimated average surface density of these features by dividing column (3) by column (4).}}
\label{tab:filament_table}
\end{table*}

\section{Dust extinction}

\begin{figure*}
	\plotone{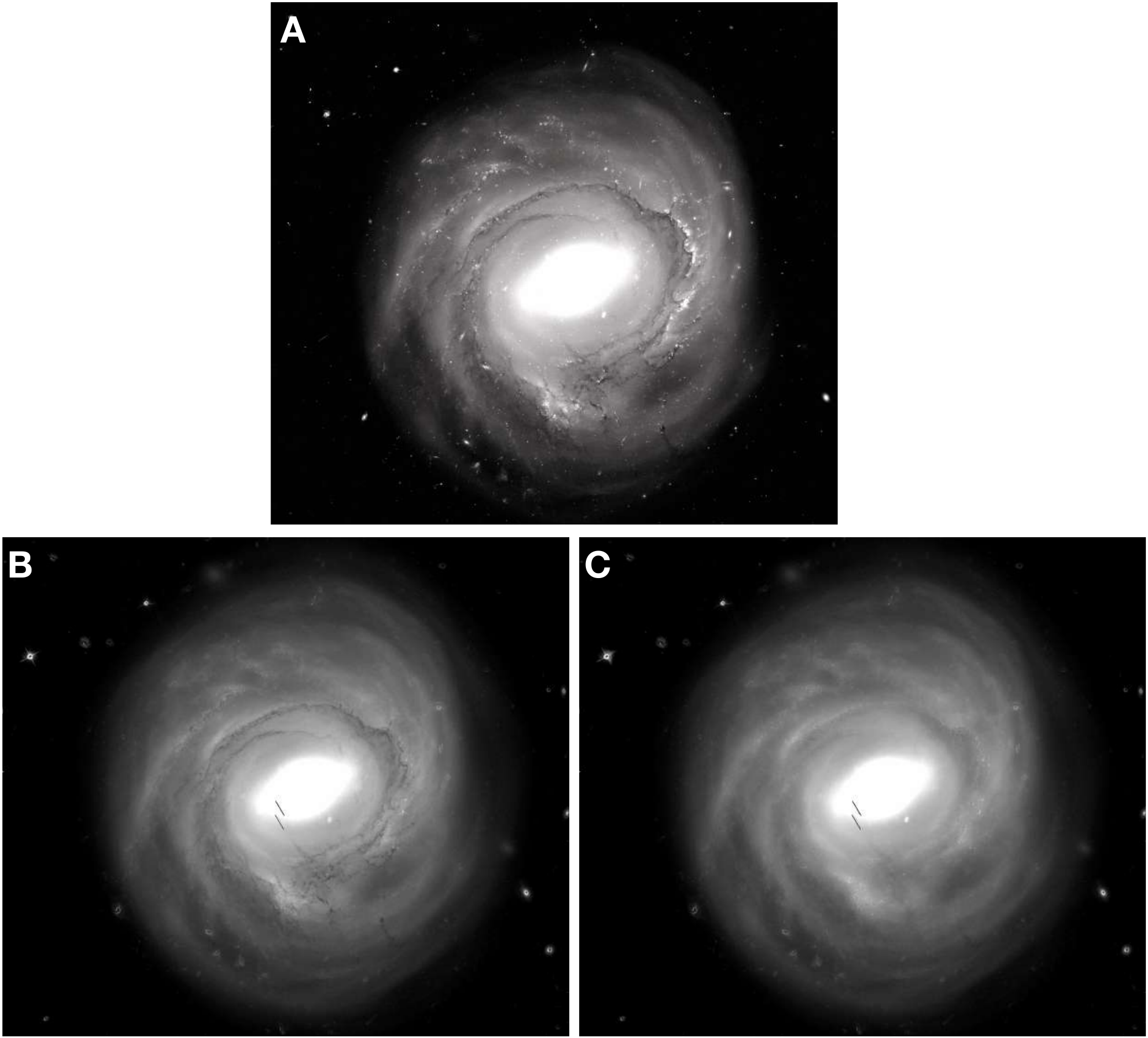}
	\caption{A progression of images showing the results of our dust extinction approximation for the full galaxy. In panel A the original F606W image is shown. In panel B, the results of the subtraction of the bright star clumps is shown. Finally, in panel C, pixels with prominent dust extinction are replaced by an estimate of their dust extinction corrected surface brightness value. Note that some large scale, low extinction dust regions may remain as it would be difficult to distinguish between these and natural surface brightness fluctuations due to arm-interam regions in a spiral galaxy like NGC 4921. Also note that in panels B \& C the black bars near the nuclear region are an artifact from the HST detector gap in this region leftover from the mosaic construction.}
	\label{fig:Dust_removal}
\end{figure*}

\begin{figure*}
    \epsscale{0.8}
	\plotone{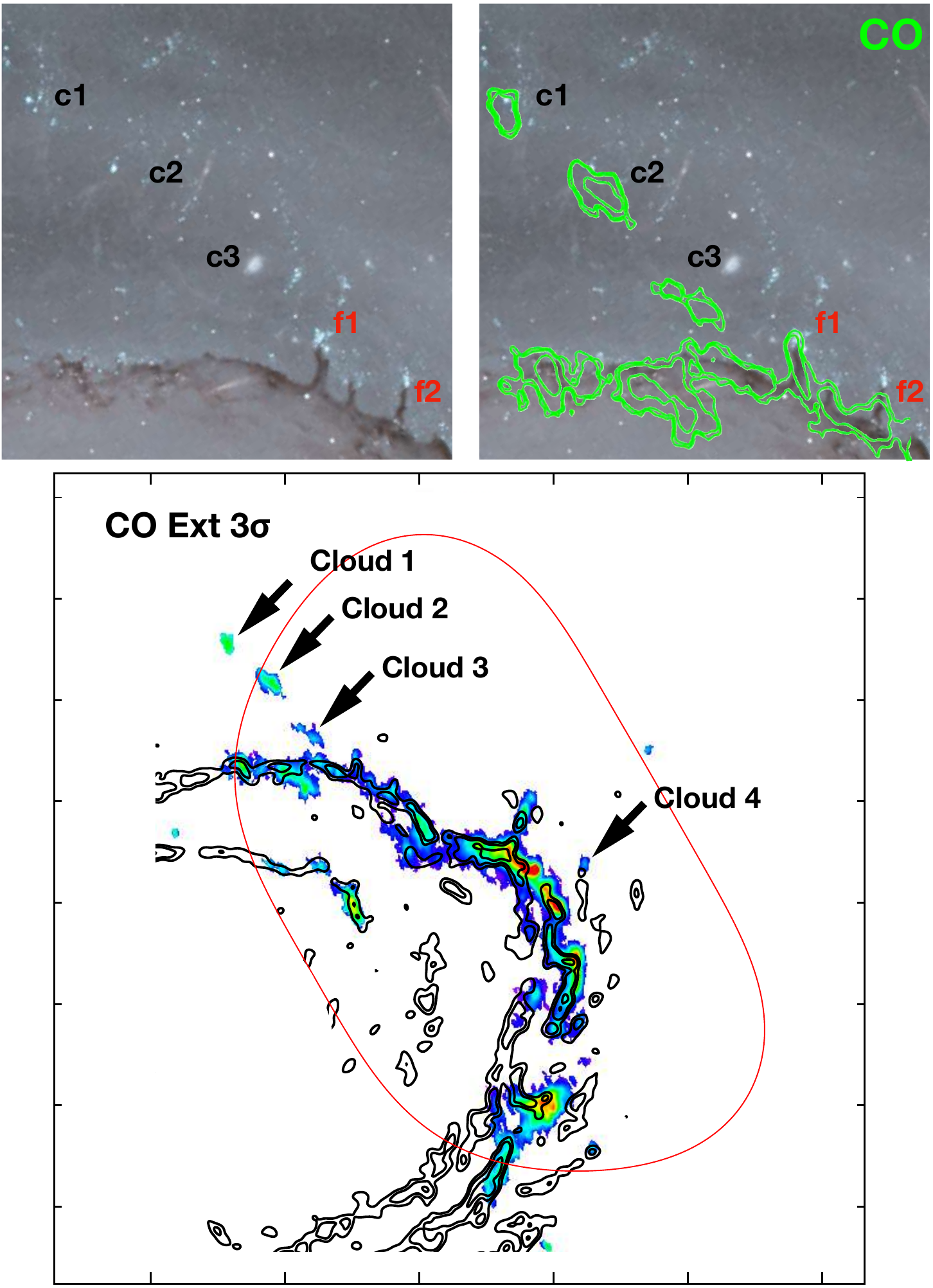}
	\caption{\textbf{Top}: A comparison of the HST color image on the left, and on the right the ALMA moment 0 map overlayed in contours on the that image. We note that the labeled filament CO gas overlaps with visible dust in the HST image, but the labeled clouds apparently do not. \textbf{Bottom}: CO intensity (moment 0) with contours from the dust extinction map overlaid. The minimum contour corresponds to a 3$\sigma$ dust extinction magnitude detection. Beyond the red polygon, the relative sensitivity of the ALMA data drops below 50\%, where it is difficult to distinguish real emission from primary beam noise. Almost all CO emission which is not located in this low sensitivity region, or in a region with a strong stellar emission clump (which makes estimating dust extinction in this region impossible) has a corresponding dust detection. The most conspicuous CO clouds which have no corresponding dust extinction detection also happen to be the CO clouds marked in Figure \ref{fig:model_residual} as Clouds $2-4$ which have notably blueshifted velocities.}
	\label{fig:dust_high}
\end{figure*}

We also study the dust extinction on the leading side of NGC 4921, to investigate its correlation with the CO(2-1) surface brightness, and in this way constrain the location of the gas and dust relative to the stars in the galaxy. We created a dust extinction map from the HST F606W filter image of the galaxy. Figure \ref{fig:Dust_removal} shows the stages of image processing for dust extinction identification described in detail as follows. First, we used the Source Extractor package from the Astromatic software suite \citep{Bertin+96} to detect surface brightness peaks in the image associated with either background galaxies or bright stellar concentrations in NGC 4921. We then flagged the pixels associated with these features, and interpolated their surface brightness value from that of nearby, non-source pixels, to create a smooth image of the disk showing only the smooth surface brightness profile of the galaxy, and the dust (Figure \ref{fig:Dust_removal}b). Then we subtracted the smooth surface brightness profile of the galaxy by using an unsharp mask image of the galaxy. Pixels in the unsharp masked image with counts below a certain threshold, chosen through trial and error to recover the most obvious dust identified visually in the HST image, were then flagged as dust extincted pixels. Using a 2D interpolation scheme, where the smooth surface brightness profile of the galaxy was fit with a 2D polynomial function, we were then able to estimate what the surface brightness of the dusty pixels should be if there were no dust extinction, with the resulting image shown in Figure \ref{fig:Dust_removal}c. This method is especially effective at identifying strong and sharp dust extinction features, but may struggle to identify large-scale, low extinction dust as it is difficult to differentiate from natural arm-interarm surface brightness fluctuations. In Figure \ref{fig:dust_high} (bottom) we show that we detect dust in nearly all pixels detected in CO(2-1), save for a few regions of interest.

\section{Discussion}
\subsection{ISM structures of particular interest}

\subsubsection{Evidence for fallback features}
\label{fallback}

Even without any image processing, it is apparent that Clouds 1, 2, and 3, to the north of the main gas ring have almost no visible dust overlapping the CO detection, especially when compared to the nearby filaments (see Figure \ref{fig:dust_high} (top)).

Clouds 2 \& 3 are also blueshifted relative to the gas disk, as seen in Figure \ref{fig:model_residual}, and appear to be the only features in the whole ALMA-HST comparison map with a clear CO detection but no dust extinction detection, as  shown in Figure \ref{fig:dust_high} bottom (regions near bright star clumps are excluded, as we cannot estimate extinction  in  them). At the azimuthal angle of these clouds, the radial velocity component becomes negligible, so the possible velocity components of the observed residual could be either $V_z$ or $V_t$. The correlation between blueshifted clouds and CO detection but no visible dust extinction suggests that these are clouds behind the midplane of the stellar disk, and are either falling back towards the galaxy at a velocity range of $\sim 20-60$ km s$^{-1}$, or rotating with significantly slower than predicted azimuthal velocity (we predict corresponding azimuthal velocity residuals of $-95$ to $-240$ km s$^{-1}$). If the clouds are behind the disk, the observer would not see any sign of dust extinction since the disk stars would be in front of the dust. Cloud 1 is likely behind the disk as it is not seen in dust, but is not falling back at this time as its velocity residual is near 0.

\begin{figure}
	\plotone{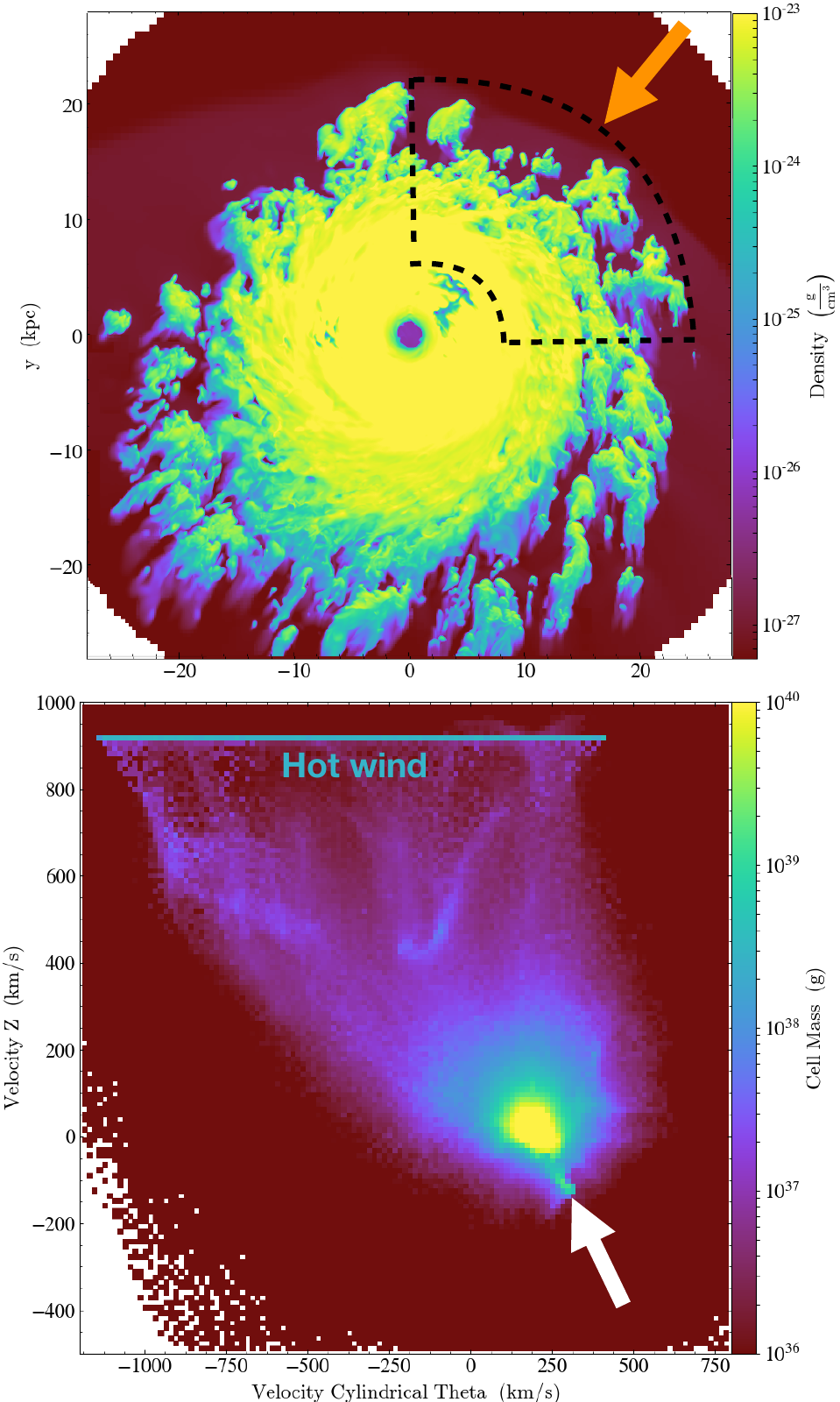}
 	\caption{\textbf{Top:} A snapshot of a simulation from \citet{Tonnesen+19} showing the maximum density of only gas $0.1-3$ kpc behind the disk for a galaxy experiencing ram pressure (at a wind angle of 53\degree). We highlight the quadrant of the galaxy equivalent to the same quadrant which we observe NGC 4921 with ALMA, rotating into the ram pressure wind front. The projected direction of ram pressure is indicated by the orange arrow. \textbf{Bottom:} A phase plot of the gas shown in the leading quadrant above, with the vertical component of the velocity on the y-axis, and the azimuthal component of the velocity on the x-axis. We indicate with the white arrow high-density material that is falling back towards the disk (has a negative $V_z$), and note that it is rotating faster than surrounding material.}
	\label{fig:phase_plot}
\end{figure}

To constrain the likelihood of the observed velocity residual being dominated by either $V_z$ or $V_t$, we consider information from a a hydrodynamical simulation that uses the adaptive mesh refinement (AMR) code ENZO and does not include magnetic fields, presented in \citet{Tonnesen+19} (run RCCW). The simulation involves a galaxy under a constant wind at a disk-wind angle of 53\degree, and a projected wind angle at 30\degree east from north, and shows gas falling back at all stages of the simulation in the same quadrant as we observed in NGC 4921. We show gas $0.1-3$ kpc behind the disk from this simulation in Figure \ref{fig:phase_plot}. We find a dense concentration of gas falling back towards the disk from behind the galaxy, similar to the decoupled clouds from our observations. In the simulation, these clouds are rotating faster, not slower, than the predicted circular velocity. This is due to the fact that, while the azimuthal component of the wind is slowing the rotation of the gas, the radial component of the wind pushes gas to smaller radius, and thus, due to conservation of angular momentum, the gas swings out to higher velocities. Because the larger component of the wind at this angle in the simulation is radial, the net effect is higher rotational velocity. The correlation between fallback and increased velocity in this quadrant persists for hundreds of megayears. This supports that the blueshifted residual velocity of the observed decoupled clouds are not in fact from an azimuthal motion, but rather from vertical motion from re-accretion.  These results are also seen in simulations by co-author Smith; as presented in \citet{Bellhouse+20}.

Re-accretion of previously stripped gas which does not reach escape velocity is predicted in simulations (e.g. \citet{Vollmer+01a, Kapferer+09, Tonnesen+10, Quilis+17}), but has not been robustly observed before. Some possible signs of fallback of stars on the disk have been postulated \citep{Hester+10, Kenney+14, Cramer+19}), but the kinematic evidence, in combination with the evidence from dust extinction that we present supporting fallback of gas is new.

\begin{table*}
\centering
\movetableright=-0.5in
\begin{tabular}{cccccc}
\hline
\head{1.5cm}{\textbf{Feature}} & \head{2.0cm}{\textbf{Integrated Flux$^{\dagger}$ (Jy km s$^{-1}$)}} & \head{2.0cm}{\textbf{Mass M$_{\mathrm{H_2}+\mathrm{He}}$ (M$_{\odot}$)}} & \head{1.5cm}{\textbf{Area$^{\dagger}$ (pc$^2$)}} & \head{2.0cm}{\textbf{Surface Density (M$_{\odot}$ pc$^{-2}$)}} \\ \hline
\textbf{Cloud 1} & 0.58 & 1.8 $\times$ 10$^7$ & 3.4 $\times$ 10$^5$ & 53 \\
\textbf{Cloud 2} & 0.93 & 2.9 $\times$ 10$^7$ & 7.2 $\times$ 10$^5$ & 40 \\
\textbf{Cloud 3} & 0.38 & 1.2 $\times$ 10$^7$ & 4.8 $\times$ 10$^5$ & 25 \\
\textbf{Cloud 4} & 0.68 & 2.1 $\times$ 10$^7$ & 3.9 $\times$ 10$^5$ & 29 \\
\end{tabular}
\caption{Properties of select CO features in NGC 4921. From left to right, in column (1): the feature name (abbreviated in some figures as `c1', `c2', and `c3'). (2): The integrated CO(2-1) flux based on the moment 0 map. (3): Total mass of H$_2$ \& He assuming $\alpha=4.3 \, \mathrm{M}_{\odot} \,\mathrm{pc}^{-2} \, (\mathrm{K} \, \mathrm{km} \, \mathrm{s}^{-1})^{-1}$ and $R_{21} \approx 0.8$, and the mass fraction of He/H$_2$ to be 1.34. (4) Surface area of features based on the spatial extent of the CO feature in dust. (5): Estimated average surface density of these features by dividing column (3) by column (4).}
\label{tab:cloud_table}
\end{table*}

\subsubsection{Properties and evolution of filament 3}
\label{filament}

Through analysis of the properties of filament 3, including the mass and velocity profile, and comparison with a similar looking filament formed in simulation, we can investigate the possible formation of these types of structures.

\begin{figure*}
	\plotone{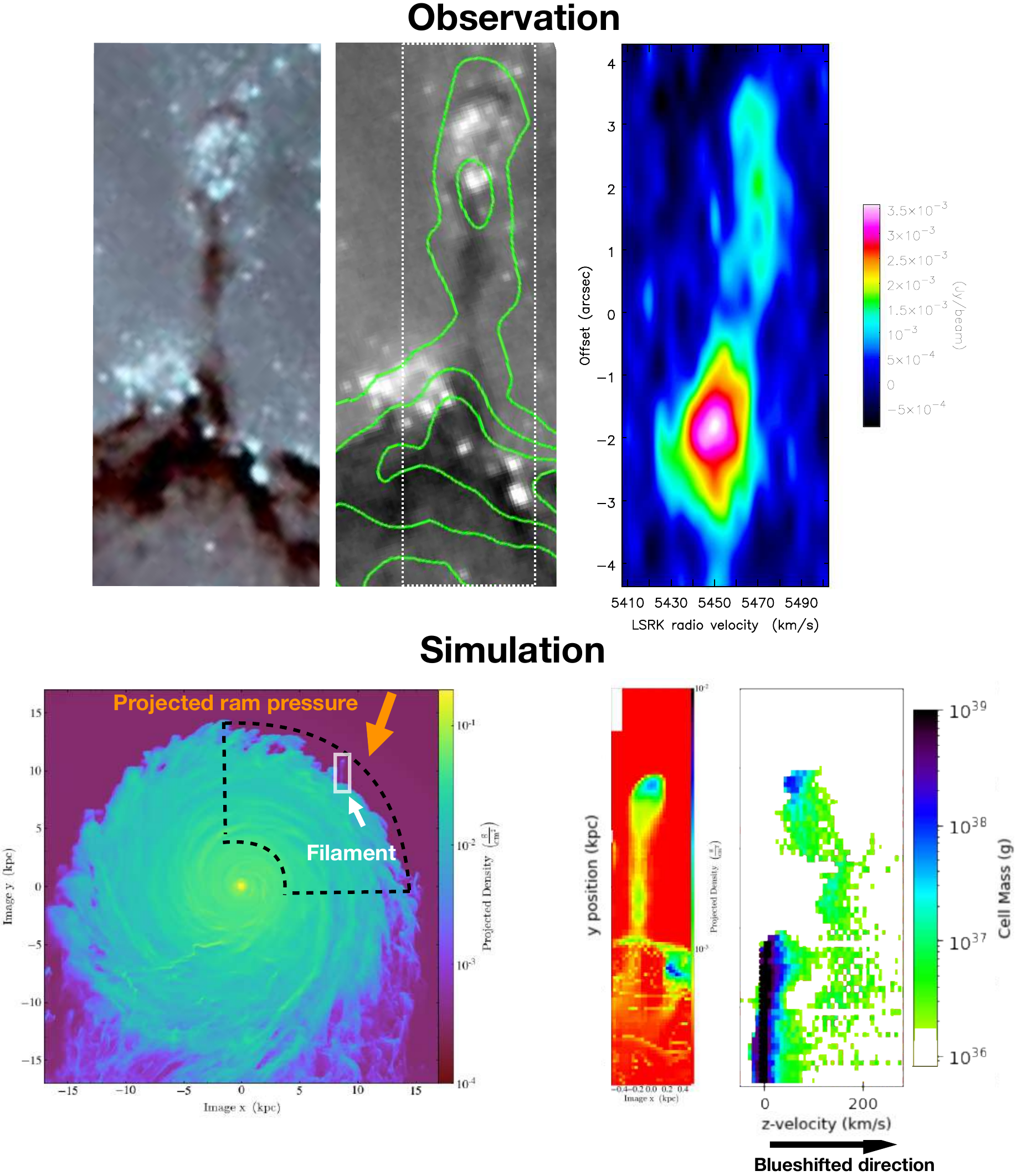}
	\caption{A zoom-in on the largest filament, labeled as Filament 3 in Figure \ref{fig:highres_zoom}. On the top row, from left to right: an HST color image, HST F606W plus CO(2-1) contours from ALMA, and a PVD along the filament, with the width and height of the PVD box indicated with the white dotted line on the left. On the bottom row, from left to right as follows. A snapshot of a simulation from \citet{Tonnesen+19} showing a galaxy experiencing ram pressure, with the projected wind angle being 30 degrees east from north. In this snapshot, we find a filamentary feature that appears similar in morphology to the filaments seen on the leading quadrant of NGC 4921. Next, a zoom in of the filament region from the simulation, restricted to the velocity range of gas around the filament so the full extent of the filament into the disk can be seen. Finally, a PVD of the zoom in box shown on the left. The black arrow highlights that, from the head to the tail of the filament, the $z$-velocity of the filament associated gas increases, the opposite of the trend seen in the filaments of NGC 4921.}
	\label{fig:big_fil}
\end{figure*}

The observed filament 3 is the largest filament we detect, with a total length of 4.5'' (2.1 kpc), and a width of $\sim$1'' (0.5 kpc). Like filaments 1 \& 2 shown previously in Figure \ref{fig:highres_zoom}, it features a complex of blue stars near the head; it also has a visible cap of dust at the top end. The fact that filament 3 is well resolved along its length (although not along its width) by our ALMA observations means that it is possible to study the mass and kinematic profile along the length of the filament. We show a zoom-in of our observations of this filament in Figure \ref{fig:big_fil} (top), including the HST image, our ALMA observations, and a PVD along the length of the filament. Overall, we detect CO along the entire length of the filament as seen in dust, including in the region of blue stars at the head where dust extinction is difficult to detect. The filament appears to be connected to the main CO ring, with a region of strong extinction in an inverted Y-shape at the base where the filament starts (see Figure \ref{fig:big_fil} top left). The filament also appears to be connected kinematically at its base to the main ring of CO. The PVD shown in Figure \ref{fig:big_fil} (top right) supports that the filament is not entirely decoupled from the surrounding ring, instead remaining connected in position-velocity space. Although the connection appears faint in position-velocity space, it is apparent in dust extinction and in the CO moment 0 map. Both the dust and CO are fainter at the connection, suggesting that this filament may be in the process of detaching from the main gas ring.

\begin{figure*}
	\plotone{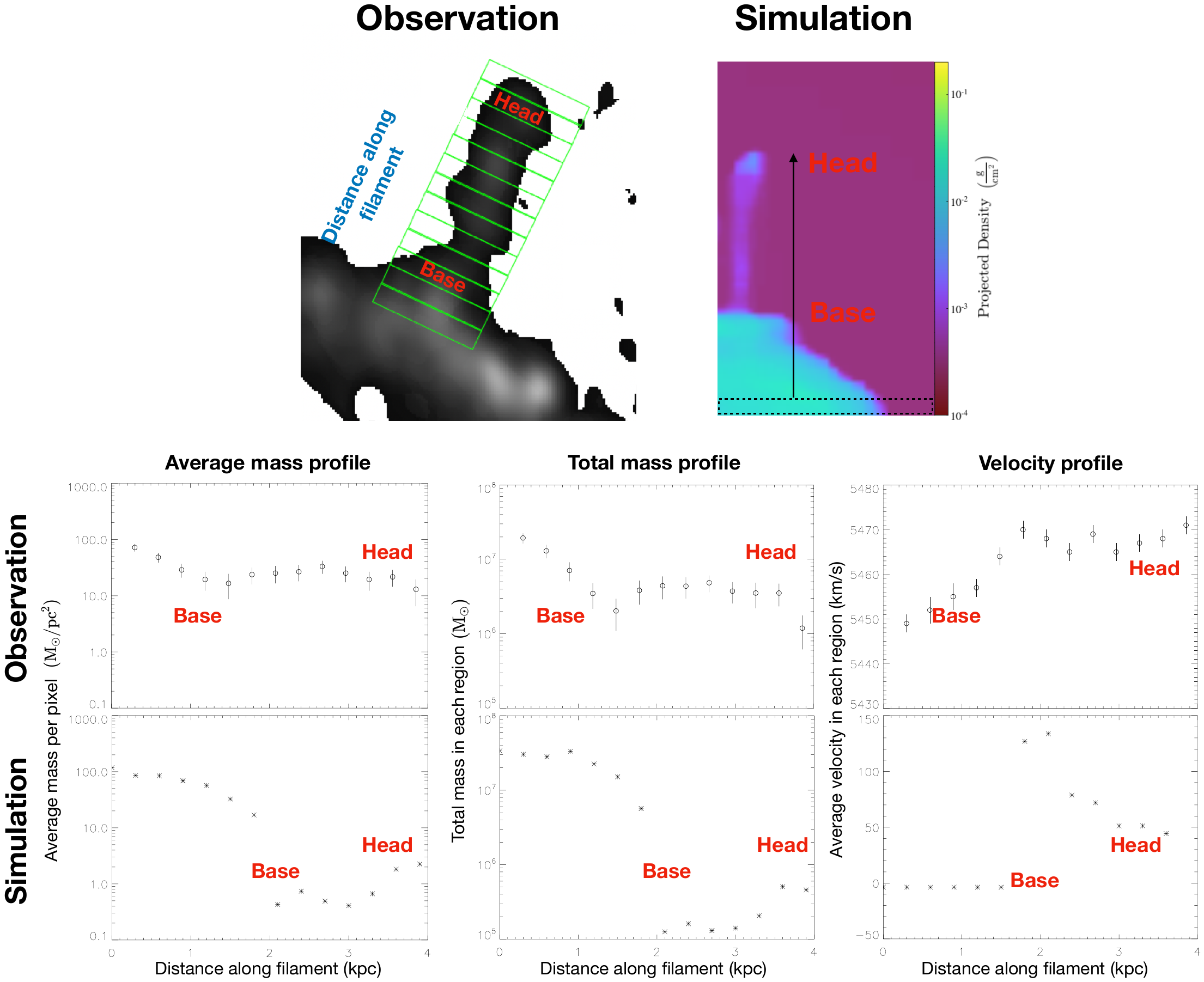}
	\caption{\textbf{Top Left}: Grey scale representation of CO(2-1) intensity with a series of 0.09 kpc$^2$ boxes along the long axis of the filament with each individual box height equal to the beam size ($\sim$ 0.45''). The region where the filament appears spatially and kinematically attached to the main ring of CO is labeled as the base, and the region where the filament overlaps with an apparently associated clump of young stars is labeled as the head. \textbf{Top Right}: The simulated filament region with the same series of 0.09 kpc$^2$ boxes shown, with the region overlapping with the disk labeled as the base, and the high projected density cloud at the tip labeled as the head. \textbf{Bottom}: At the left, the average molecular gas mass in a detected pixel in each box region in M$_{\odot}$ pc$^{-2}$. This is shown due to the fact that the width of the filament is not uniform along its length, so simply totalling the mass in each box may not give the complete picture. At the middle, The total mass of molecular gas in each box region in M$_{\odot}$. At the right, the velocity profile, measured as the surface brightness weighted average velocity in each box region.
	}
	\label{fig:big_fil_mass}
\end{figure*}

To further investigate the spatial and kinematic profile of filament 3, we plot the total mass, average mass, and velocity profile in a series of rectangular regions along the filament, shown at the top of Figure \ref{fig:big_fil_mass}. We find for the observed filament that just as is seen in dust, the region just past the base of the filament has a lower total mass of gas than either the base or further along the filament. The region further along the filament from its base is fairly uniform in mass per length, and has relatively constant velocity. However, there remains a significant velocity offset ($\sim$ 20 km s$^{-1}$ in the redshifted direction compared to the predicted circular velocity from our model) between the gas along the body and head of the filament and the attached ring. The head of the filament may have been more dense in the past before the onset of star formation as some gas originally in the head has been converted to stars. Furthermore the effects of feedback after star formation may have already blown away some of the molecular gas. 

The formation of this type of filament, and the influences that affect its shape and evolution, are still not entirely clear. One possible formation scenario we considered is that the head of the filament is a dense region of gas which resisted ram pressure as the surrounding gas was pushed away, and that the body of the filament is ablated gas that was stripped away from the dense cloud at the head. This is seen to happen in hydrodynamical simulations, such as those by \citet{Tonnesen+12},
which use the adaptive mesh refinement (AMR) code ENZO, although do not include magnetic fields. We show an example of this type of filament feature in Figure \ref{fig:big_fil} (bottom). In the simulation snapshot shown, there is a high projected density gas cloud beyond the stripping radius, with a tail of material streaming off it that overlaps in projection with the downstream disk. However, the PVD of the simulated filament (Figure \ref{fig:big_fil}), as well as the mass and velocity profile along the simulated filament (Figure \ref{fig:big_fil_mass}), are quite different from those of the observed Filament 3. While the body of the simulation filament is also offset in the ram pressure direction (blueshifted), in contrast to filament 3 the further from the head of the simulated filament, the more ram pressure appears to have accelerated the gas. In the region of projected overlap with the disk, the simulated filament does not appear kinematically connected to the disk gas, but is instead offset from the disk gas by $\sim 150-200$ km s$^{-1}$.

The PVD, and velocity profile of the filament in the simulation is consistent with this filamentary feature being formed from ablated gas originating from the cloud at the head, which is highly accelerated by ram pressure, such that we see a clear positive velocity gradient along the filament from head to base in the direction of ram pressure. In contrast, in the observed filament, we do not see this sort of profile, instead the gas along the filament beyond the base region has relatively constant velocity. Moreover, at the base of the filament, it appears connected both spatially and kinematically to the surrounding disk gas.

Furthermore, the mass of the ablated cloud in the simulation is concentrated at the head, and lower along its length. It is also much lower than the region where it overlaps with the disk when compared to filament 3 (Figure \ref{fig:big_fil_mass}). The observed mass and velocity profile support that the filaments we observe in NGC 4921 are held together by magnetic binding that inhibits decoupling of low and high density gas in the filament, as the region undergoes stripping. This scenario was first suggested by \citet{Kenney+15} based on the dust morphology. Most detailed simulations of ram pressure stripping do not include star formation and/or magnetic fields in both the galaxy and the ICM. Our observations support that a more complete understanding of these factors, especially on the leading side of ram pressure affected galaxies, are necessary to fully understand the formation of structures like filament 3.

Finally, we consider the implication of the observed redshift when compared to the circular velocity model along the observed filament. First we consider whether the observed redshift could be consistent with the filament being formed via radial pushing of the surrounding gas. As these filaments are only observed in the leading quadrant between the area of the main CO ring of PA $\sim$ 25\degree \, and 55\degree \, from the major axis, we assume they formed within the time it took the galaxy to rotate between 25\degree and 55\degree. The circular velocity at the radius of the filament ($\sim$ 30'' or 14.8 kpc) as determined from DiskFit and the Tully-Fisher relation is $\sim$ 300 km s$^{-1}$, thus we estimate the rotation period of the disk here is $\sim$ 300 Myr. The fractional time the galaxy would take to rotate from 25\degree \, and 55\degree \, is thus $\sim$ 30 Myr. If we consider a scenario in which the head of the filament was originally part of the main ring, and that the main ring was then pushed down to where it lies now, i.e. moved the distance of the length of the filament, that would be a total distance traveled of $\sim$ 2 kpc. The gas pushed by ram pressure would need to travel at an average velocity, in the radial direction of $\sim$ 65 km s$^{-1}$ to match this distance traveled over 30 Myr (assuming for estimation that ram pressure is constant over this time period). We previously calculated that, when compared to the kinematic model of the circular velocity, filament 3 has a velocity residual of $\sim 20$ km s$^{-1}$, which could be due to a radially inward velocity of up to 95 km s$^{-1}$, although any of the three components of motion could contribute (as described in Equation \ref{velocity_eq}). A radially inward motion for the filament is difficult to explain given that the filament appears to be the only gas remaining in the otherwise stripped zone; we would expect the downstream gas to have a strong radial motion residual, but not the filament. We still do not fully understand the kinematics of these filaments. Uncertainty inherent in our approach to generating the circular velocity model is also a source of systematic error that makes interpreting these residuals more difficult. We explore these uncertainties further in Section \ref{evolution}.

\subsubsection{C-shaped pockets}
\label{c-shape}

In \citet{Kenney+15}, the authors identified several c-shaped dust features along the leading edge of the gas ring in the NW and SW quadrants of the galaxy. While we do not have sufficient resolution to study all of the c-shapes they identified, we do resolve one particular c-shaped pocket starting at an azimuthal angle $\sim$ 65\degree W from N, slightly downstream of the large filament, shown in Figure \ref{fig:filament_pvds}. Like filament 3, we cannot establish with absolute certainty the relative contribution of each component of motion from Equation \ref{velocity_eq} to the overall observed velocity residual when compared to the model of circular motion, but we can make some deductions based on geometry. First, this feature is near the minor axis so contributions from azimuthal motion will be minimal. Second, since the c-shapes are on the far side of the disk, a radially inward motion at the C-shape will result in a blueshift to the observer. Finally, the vertical component of ram pressure is mostly in the redshifted direction. Therefore the net velocity shift depends on the relative strength of the radial and vertical components. Since the cup of the c-shape is relatively blueshifted with respect to the edges, this is consistent with an inward radial motion stronger than the vertical component of the ram pressure. The velocity offset between the edges and the cup is $\sim$ 10 km s$^{-1}$, which, assuming the motion is entirely radial, corresponds to $V_{r}=30$ km s$^{-1}$. The kinematic pattern of this feature can be understood if the gas at the edges of the `c' is not radially pushed as strongly as was presumably lower density gas at the center of the `c'. The morphology of this feature appears to be consistent with this radial pushing scenario as well. The moment 0 map of the c-shape shows that the tips of the `c' are much denser than the cup of the c-shaped feature. The gas distribution may be due to the cup of the c-shape being an interarm region where the ISM is less dense than the arms. This type of stripping process may be related to observations that ram pressure can unwind the spiral structure of a galaxy \citet{Bellhouse+20}, although further study is needed to confirm.

Furthermore, magnetic fields are likely a significant factor in forming the structure of the c-shape by inhibiting decoupling, keeping the gas linked and forming the smoothly curved c-shape \citep{Kenney+15}.

\subsection{Estimated RPS around NGC 4921}
\label{RPS_estimate}

We have estimated the surface density of a number of ISM features that appear to be affected by ram pressure. To analyze the role of ram pressure in the formation and evolution of these features, we would like to estimate the strength of ram pressure and the opposing restoring force from the disk around the affected region in NGC 4921. By estimating the ICM gas density $\rho_{I C M}$, the relative velocity of the galaxy with respect to the cluster $v$, and the restoring force per unit mass $d\Phi_g/dz$, we can use the \citet{Gunn+72} stripping criteria $\rho_{I C M} v^{2} \geqslant \Sigma_{I S M} \frac{d \Phi}{d z}$  to estimate the surface density of gas $\Sigma_{I S M}$ that would be stripped. We estimate $\rho_{I C M}$ to be $\sim$ $7 \times 10^{-28}$ g cm$^{-3}$, based on a $\beta$-profile for an isothermal spherical distribution with coefficients given by \citet{Mohr+99, Fossati+12}, and the 3D distance of NGC 4921 to cluster center being equal to its projected distance of $\sim$675 kpc. The line of sight velocity of NGC 4921 with respect to the cluster center, estimated as the average of the line of sight velocity of the two brightest central elliptical galaxies in Coma, NGC 4889 and NGC 4874, is $v_{\rm gal}-v_{\rm Coma} = 1500$ km s$^{-1}$. The leading side of NGC 4921 has a significantly different dust morphology than the trailing side, being much more compressed due to the influence of a component of ram pressure acting in the plane of the sky. Thus, the true 3D velocity of NGC 4921 is likely to be larger than the line of sight velocity. We estimate the maximum 3D velocity according to simulations for galaxies in a Coma-like cluster from \citet{Jachym+17}, that found an infalling galaxy with an open orbit would have a 3D velocity of at least $\sim 3000$ km s$^{-1}$. Thus the maximum ram pressure we estimate for NGC 4921 is approximately $P_{\mathrm{ram}}=\rho v^2 \sim 6 \times 10^{-11}$ dyne cm$^{-2}$.

We estimate the gravitational restoring force per unit mass $\frac{d \Phi}{d z}$ as approximately equal to $V^2_{rot} R^{-1}$, as done in \citet{Cramer+20}. At the radial distance of the outer edge of the main dust ridge, $r\sim 30$''$\, = 18$ kpc, the circular velocity predicted from Tully-Fisher is $\sim$ 300 km s$^{-1}$, so $V^2_{rot} R^{-1} = 2 \times 10^{-13}$ km s$^{-2}$. Using the Gunn \& Gott formula, we estimate the surface density of gas that would be stripped under these conditions as $\Sigma_{I S M} \approx 18$ M$_{\odot}$ pc$^{-2}$ for the upper limit velocity estimate.

\subsection{Evolution of gas in the leading quadrant}
\label{evolution}

Both the estimated surface density of gas that could be stripped by ram pressure, and the surface density of features in the ISM we have estimated in this paper, have significant sources of uncertainty, like the true 3D velocity of the galaxy, and the X factor for converting CO to H$_2$ mass $\alpha_{\mathrm{CO}}$. Despite these uncertainties, our estimates of the surface density of gas that could be susceptible to stripping are consistent with the observed properties of the stripped zone. The decoupled clouds of gas detailed in Table \ref{tab:cloud_table} have estimated surface densities between $\sim25-55$ M$_{\odot}$ pc$^{-2}$ (measured at a resolution of $220 \times 170$ pc); for them to be falling back onto the galaxy we would indeed expect them to be denser than the maximum surface density of gas that could be stripped by our estimate of the ram pressure. Furthermore, the filaments are the only other ISM surviving in the otherwise stripped zone beyond $r\sim 30$'', and they also have estimated surface densities higher than that which we expect to be stripped (shown in Table \ref{tab:filament_table}) of $\sim26-44$ M$_{\odot}$ pc$^{-2}$. Interestingly, a very similar filament feature with star formation at the head seen in NGC 4402, a ram pressure stripped galaxy in the Virgo cluster, was found to have an estimated surface density of only 7 M$_{\odot}$ pc$^{-2}$ \citep{Cramer+20}. The Virgo cluster has a much lower total mass than the Coma cluster, and generally has much weaker ram pressure. It would be interesting if the surface density of these unique filamentary features seen in ram pressure affected galaxies were related to the conditions of the ram pressure the respective galaxies were experiencing. However, a larger sample of these types of features would be needed to establish this sort of connection.

Overall, the measured surface density of NGC 4921 ISM features remaining in the otherwise stripped zone is consistent with all lower surface density gas having been stripped leaving behind only these dense clouds. This stripping of gas could occur from pushing of the low density gas, or evaporation of the low density gas. It is interesting that overall we detect no large scale significant velocity residual on the leading side; on the surface this seems to support the evaporation scenario. However, there is more evidence supporting that gas has been radially pushed by ram pressure. First,  in a similar kinematic analysis of molecular gas in the ram pressure stripped galaxy NGC 4402 in the Virgo cluster, \citet{Cramer+20} found that the leading side of the disk of that galaxy had velocity residuals of up to 60 km s$^{-1}$, consistent with gas being pushed by ram pressure. Our CARMA observations of NGC 4921, as well as previously taken HI observations, also appear to be consistent with gas having been radially pushed by ram pressure. As noted previously, the leading quadrant has the highest peak surface brightness in the CARMA CO map. The HI contours in this region are also more compressed than anywhere else in the galaxy. It was also noted that in the leading quadrant, there were a much higher number of young star concentrations than anywhere else in the galaxy \citep{Lee+16}. This is consistent with triggering of star formation from gas compression on the leading side. Furthermore, it can be seen in Figure \ref{fig:highres_zoom} that on the leading side, young stars and molecular gas are spatially offset, with molecular gas downstream from the young stars. This supports that star formation has been triggered earlier in the ram pressure stripping process, and gas has since been pushed inward radially from where this star formation took place. Finally, the c-shaped features have higher velocity residuals in the cup than at the edges, consistent with lower density gas in the cup being pushed further radially inward than denser gas at the edges.

There are a number of factors which could explain why, despite evidence for radial pushing of gas by ram pressure in NGC 4921, we do not observe a large scale velocity residual like we do in NGC 4402. We may not be able to detect any velocity residual in NGC 4921, even if the gas is being significantly kinematically affected, due to projection and the ram pressure disk-wind angle. If the disk-wind angle is close to edge-on, most of the velocity residual will not be along the line of sight. 

Furthermore, we note that our approach to estimating the rotation curve via a kinematic model of the molecular gas that assumes only normal circular motion is limited by the molecular gas in the galaxy likely being disturbed by ram pressure (as well as bar and spiral arm motions). We believe this is mitigated due to the fact that the kinematics of the gas on the leading side are likely to be more strongly influenced by ram pressure (as was seen in \citet{Cramer+20}), and thus by excluding this region from the fit, the residuals in this region based on a fit to the rest of the galaxy should still reveal some elevated level of residuals on the leading side. However, significant uncertainty still remains; for example, a different rotation curve that is fit to reduce the N-S residuals would result in different residuals in the area of the leading side we study. Ultimately, observations of stellar kinematics are likely necessary to make more accurate measurements of the non-circular motions of the gas, as the stellar kinematics should not be strongly affected by ram pressure.

Keeping this significant uncertainty in mind, if the model is indeed close to accurate, there are number of explanations for the velocity residual patterns we observe. First, we note that we will only observe a velocity difference between the different gas densities if they are actively in the process of being pushed back. Whereas if all the gas that could be pushed already had been, we would not observe a residual. It is possible that the disk-wind angle in NGC 4921 contributes to gas being pushed radially into already dense gas in the main ring of NGC 4921. This ring of gas is seen in dust along nearly the entire galaxy. Gas pushed into this ring could slow down as it was compacted against this dense, more strongly gravitationally bound gas front, leading to an overall lower velocity residual than gas on the leading side of NGC 4402 which is being pushed in the direction out of the disk, instead of into it.

\section{Conclusion}

We have presented new CO observations of the massive nearly face-on spiral galaxy NGC 4921 in the Coma cluster. These include CARMA observations of CO(1-0) covering the whole galaxy (with a beam size of 6.5'' $\times$ 4.8'', or 3.2 $\times$ 2.4 kpc), and high resolution ALMA CO(2-1) observations of the leading quadrant (with a beam size of 0.45'' $\times$ 0.35'', or 220 $\times$ 170 pc). The effects of active ram pressure stripping are identified in the morphology, kinematics, and properties, such as mass, size, and surface density, of features detected in CO.

\begin{description}
	\item[$\bullet$ Compression of molecular gas] We find evidence from CARMA observations of the whole galaxy that the molecular gas distribution is concentrated on the leading quadrant of the galaxy, which is experiencing the strongest ram pressure. The overall peak surface brightness in CO(1-0) is located near the azimuthal angle at which the ram pressure is likely to be incident, suggesting molecular gas is being compressed here. This compressed region of molecular gas is radially downstream from visually apparent concentrations of young stars, suggesting that the gas that formed those stars has continued to be pushed radially inward. This is direct evidence that the observed elevated star formation rates in the disks of actively ram pressure stripped galaxies is due to gas compression.
	
	\item[$\bullet$ Molecular gas in filaments] Several large (ranging from $0.5-2$ kpc long) filaments of dust and molecular gas are found extending into the region of the leading quadrant that otherwise appears nearly completely stripped of gas and dust. The largest of these filaments is observed to have clumps of young stars at the head, and all three filaments have molecular gas and dust along their length. These types of filaments are also found exclusively in a narrow azimuthal range on the leading side suggesting they form in this zone and have an overall short lifetime. They have average surface densities ranging from $25-45$ M$_{\odot}$ pc$^{-2}$ measured at $220 \times 170$ pc resolution. 
	
	One filament we detect is especially large, with a length of 2 kpc, a width of 0.5 kpc, and a molecular gas mass of 4.3 $\times$ 10$^7$ M$_{\odot}$, corresponding to an average surface density of 44 M$_{\odot}$ pc$^{-2}$. The size of this filament, and the resolution of our ALMA data, allows us to study the mass and kinematic profile along the length of the filament. While we find the average surface density of the filament drops close to where it attaches to the ring, overall the filament has relatively uniform mass distribution along its length. 
	
	We find striking differences between the observed properties of the largest filament with those of filaments formed from ablated decoupled clouds seen in AMR hydrodynamical simulations of ram pressure stripping events. The base of the observed filament appears both spatially and kinematically connected to the main ring of molecular gas, whereas the simulated ablated filament is significantly offset in velocity from the main gas ring where they spatially overlap. This suggests that the observed filament is truly connected to the main gas ring, but the simulated filament is not. Moreover, the mass profiles are very different, with the observed filament having a much higher gas surface density than the simulated one. We conclude that the observed filaments are not formed simply from decoupled clouds with ablated tails, since they are ‘stickier’ and ‘heavier’ than the filaments in hydrodynamical simulations. The formation of the observed filaments must involve physics that is missing from the simulations, such as magnetic fields, and we propose that the observed filaments form through cloud decoupling inhibited by magnetic binding.
    
    \item[$\bullet$ Fallback clouds] Fallback of stripped gas which does not reach escape velocity has been predicted in a number of simulations (e.g. \citet{Vollmer+01, Roediger+07, Jachym+09, Tonnesen+12}). We have found the first observational evidence for the fallback of some molecular clouds, on the leading side of NGC 4921. We find three clouds with blueshifted velocities ranging from $25-50$ km s$^{-1}$ and surface densities estimated to range from $25-55$ M$_{\odot}$ pc$^{-2}$. These clouds are also the only prominent CO detections which have no observed dust extinction counterpart, with an CO emission to dust extinction ratio at least 10 times higher than that in the main CO ring. This supports our hypothesis that the clouds are behind the galaxy disk midplane and falling back towards it. An analysis of fallback clouds in simulations provides evidence that the blueshifted velocity residuals of these clouds are due to vertical fallback and not slower azimuthal motion.

\end{description}

\acknowledgments

We thank the referee for their thorough reading of this paper and very helpful comments and suggestions. We gratefully acknowledge support from the NRAO Student Observing Award Program, under award number SOSPA4-003, and from the Joint ALMA Observatory Visitor Program. This paper makes use of the following ALMA data: ADS/JAO.ALMA\#2016.1.00931.S. ALMA is a partnership of ESO (representing its member states), NSF (USA) and NINS (Japan), together with NRC (Canada), MOST and ASIAA (Taiwan), and KASI (Republic of Korea), in cooperation with the Republic of Chile. The Joint ALMA Observatory is operated by ESO, AUI/NRAO and NAOJ. The National Radio Astronomy Observatory is a facility of the National Science Foundation operated under cooperative agreement by Associated Universities, Inc. Based on observations made with the NASA/ESA Hubble Space Telescope, and obtained from the Hubble Legacy Archive, which is a collaboration between the Space Telescope Science Institute (STScI/NASA), the Space Telescope European Coordinating Facility (ST-ECF/ESA) and the Canadian Astronomy Data Centre (CADC/NRC/CSA). 

T.W. acknowledges support from NSF through grants AST-1139950 and AST-1616199. Support for CARMA construction was derived from the Gordon and Betty Moore Foundation, the Eileen and Kenneth Norris Foundation, the Caltech Associates, the states of California, Illinois, and Maryland, and the NSF. CARMA development and operations were supported by funding from NSF and the CARMA partner universities.

\software{CASA (v5.4.0; \citet{McMullin+07}), Python maskmoment (\url{https://github.com/tonywong94/maskmoment}) IDL mommaps (\url{https://github.com/tonywong94/idl\_mommaps}), MIRIAD \citep{Sault+95})}

\clearpage

\bibliography{Bibliography}{}
\bibliographystyle{aasjournal}

\end{document}